\documentclass[manuscript]{acmart}
\usepackage{dirtytalk}

\usepackage{graphicx}
\usepackage{subfigure}
\usepackage{multirow}
\usepackage{calc}

\usepackage{textcomp}
\usepackage{siunitx}
\usepackage{soul,xcolor}
\usepackage{gensymb}
\usepackage{tabularx}

\AtBeginDocument{%
  \providecommand\BibTeX{{%
    \normalfont B\kern-0.5em{\scshape i\kern-0.25em b}\kern-0.8em\TeX}}}

\setcopyright{acmcopyright}
\copyrightyear{2018}
\acmYear{2018}
\acmDOI{10.1145/1122445.1122456}

\acmConference[Woodstock '18]{Woodstock '18: ACM Symposium on Neural
  Gaze Detection}{June 03--05, 2018}{Woodstock, NY}
\acmBooktitle{Woodstock '18: ACM Symposium on Neural Gaze Detection,
  June 03--05, 2018, Woodstock, NY}
\acmPrice{15.00}
\acmISBN{978-1-4503-XXXX-X/18/06}

\begin{document}

\setstcolor{red}


\title[Visualizing Human-AI Collaboration Disclosures]{More Human or More AI? Visualizing Human-AI Collaboration Disclosures in Journalistic News Production}


\author{Amber Kusters}
\orcid{0009-0004-7895-7539}
\affiliation{%
  \institution{Centrum Wiskunde \& Informatica}
  \city{Amsterdam}
  \country{the Netherlands}
}
\author{Pooja Prajod}
\orcid{0000-0002-3168-3508}
\email{Pooja.Prajod@cwi.nl}
\affiliation{%
  \institution{Centrum Wiskunde \& Informatica}
  \city{Amsterdam}
  \country{the Netherlands}
}
\author{Pablo Cesar}
\orcid{0000-0003-1752-6837}
\affiliation{%
  \institution{Centrum Wiskunde \& Informatica and}
  \institution{TU Delft}
  \country{The Netherlands}
  }
\author{Abdallah El Ali}
\orcid{0000-0002-9954-4088}
\affiliation{%
  \institution{Centrum Wiskunde \& Informatica and}
  \institution{Utrecht University}
  \country{The Netherlands}
  }


\renewcommand{\shortauthors}{Preprint - Accepted to ACM CHI 2026}

\begin{abstract}

Within journalistic editorial processes, disclosing AI usage is currently limited to simplistic labels, which misses the nuance of how humans and AI collaborated on a news article. Through co-design sessions (N=10), we elicited 69 disclosure designs and implemented four prototypes that visually disclose human–AI collaboration in journalism. We then ran a within-subjects lab study (N=32) to examine how disclosure visualizations (Textual, Role-based Timeline, Task-based Timeline, Chatbot) and collaboration ratios (Primarily Human vs. Primarily AI) influenced visualization perceptions, gaze patterns, and post-experience responses. We found that textual disclosures were least effective in communicating human-AI collaboration, whereas Chatbot offered the most in-depth information. Furthermore, while role-based timelines amplified AI contribution in primarily human articles, task-based timeline shifted perceptions toward human involvement in primarily AI articles. We contribute Human-AI collaboration disclosure visualizations and their evaluation, and cautionary considerations on how visualizations can alter perceptions of AI’s actual role during news article creation.

\end{abstract}

\begin{teaserfigure}
       \centering
       \setlength{\abovecaptionskip}{2pt}
       \includegraphics[width=0.9\textwidth]{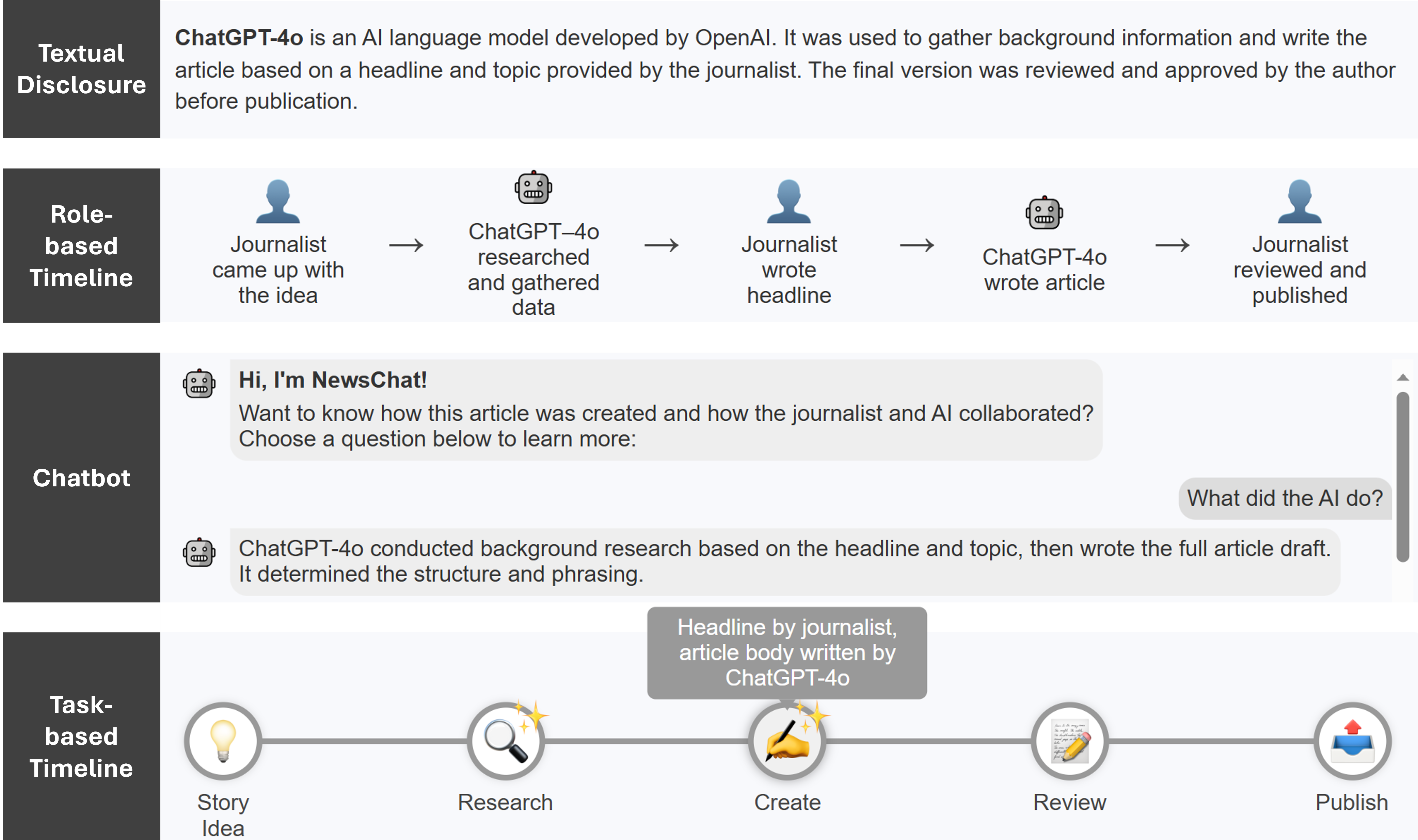}
               \caption{Our human–AI collaboration disclosure visualizations, with an example for a primarily AI-written article. From top to bottom: (1) Textual Disclosure, (2) Role-based Timeline, (3) Chatbot interface, and (4) Task-based Timeline.}                
       \label{fig:teaser}
       \Description{Four images showing the human–AI collaboration disclosure visualizations prototypes, with an example for a primarily AI-written article. From top to bottom: (1) Textual Disclosure: describing the used AI model, its contributions, and the journalist’s contributions, (2) Role-based Timeline: icons showing linear progression of the roles between the journalist and AI, (3) Chatbot: menu-based with predefined question, and (4) Task-based Timeline: icons of five stages of news article creation, hover for more information.} 
   \end{teaserfigure}



\maketitle

\section{Introduction}






Artificial intelligence (AI) technologies such as large language models (LLMs) are increasingly being used for a variety of tasks in the newsroom \cite{nishal2024envisioning}. Generative AI (GenAI) tools like ChatGPT, Gemini, and Midjourney can produce high-quality multimodal outputs based on user inputs \cite{epstein2023art, li2024user}. These developments have the potential to affect both users and the media ecosystem, essentially blurring the lines between fiction and reality as users consume news media content. Already in 2024, model output quality has advanced to a point where human observers can no longer reliably distinguish between generative AI (GenAI) output and that produced by humans \cite{groh2023human}. This further intensified with recent work on Turing tests demonstrating near-chance human detection rates of state-of-the-art LLMs, where some instances were judged to be more human than real human interlocutors \cite{jones2025largelanguagemodelspass}. These tools are often used in collaborative workflows where GenAI assists journalists rather than replacing them entirely. Prior work has shown that journalists use GenAI for tasks such as news generation, validation, distribution, and moderation \cite{cools2024uses}, and primarily for content production, particularly when it comes to text-based outputs \cite{diakopoulos2024generative}. This is also in line with the global public perception of AI usage in journalism, which includes text editing, translation, and data analysis, but also with some believing that the entire article has been written by AI \cite{fletcher2024does}.

As GenAI use in journalism grows, ethical concerns regarding transparency are rising, centering around trust and misinformation \cite{forja2024ethical, porlezza2024ai}, as well as privacy and accountability \cite{sonni2024digital}. This is further pronounced considering that current algorithmic techniques for detecting text (co-)produced by GenAI systems (e.g., ChatGPT) remain unreliable \cite{sadasivan2025aigeneratedtextreliablydetected}, similarly with cryptographic approaches that aim to encode provenance in digital media, for example in images (e.g., Coalition for Content Provenance and Authenticity \cite{c2paIntroducingOfficial}) or audio (e.g., SynthID \cite{SynthID}). Even if such detection techniques were flawless, how provenance signals should be disclosed or communicated to end users remains a key challenge \cite{el2024transparent}. Recent studies found that most readers (sampled from six countries) expressed a desire for disclosure when exposed to journalistic tasks that are produced using `AI with some human oversight' \cite{fletcher2024does}.  Moreover, regulatory frameworks such as the AI Disclosure Act of 2023 in the United States \cite{Torres2023} and Article 50 of the EU AI Act \cite{euaiact2024} articulate requirements on disclosing AI usage, which is intended to mitigate these risks. However, the regulation remains vague with terms such as "authentic" and "context of use" which leaves room for interpretation, creating challenges in design and implementation \cite{el2024transparent}. Additionally, regulation may be too limited, exempting disclosure after human review even though audiences still expect to be informed of AI usage \cite{piasecki2024ai}, in addition to expecting standardized measures and sourced references \cite{morosoli2025public,Venkatraj2025}. While different types of disclosures have been researched, from certification labels \cite{scharowski2023certification}, to detailed AI Usage Cards \cite{wahle2023ai}, or showing more details progressively \cite{springer2020progressive}, it remains a core challenge of how to not only disclose that AI was used, but also to capture the nuance of how humans and AI collaborated. Human-AI collaboration refers to a dynamic and adaptive partnership in which humans and GenAI systems work together towards a shared goal \cite{saha2023human, fragiadakis2024evaluating, holter2024deconstructing}. Representing and disclosing how much of the content was created by humans versus GenAI is important as GenAI usage can alter journalists' sense of authorship \cite{hwang202480}, and varies across writing stages and tasks. Yet, prior work has not explored how to visually represent these collaboration ratios across editorial stages within the journalistic workflow.

\begin{figure}
    \centering
    \includegraphics[width=0.6\linewidth]{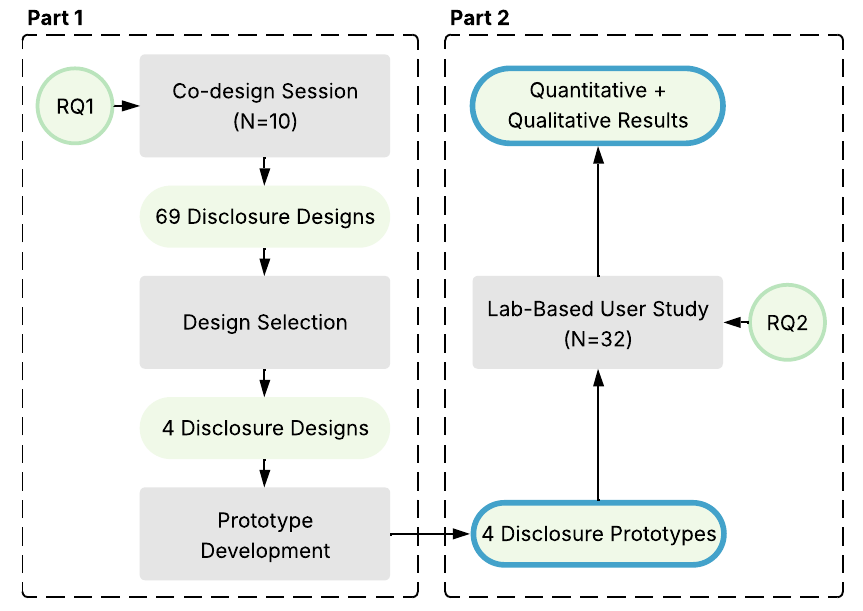}
    \caption{The two part study approach with contributions outlined in (bold) blue.}
    \label{fig:studyApproach}
    \Description{Diagram of a two-part study approach linking design generation to evaluation. Part 1 shows Research Question 1 leading to a co-design session with 10 participants, resulting in 69 disclosure designs. After design selection, four disclosure designs were developed into prototypes. Part 2 shows these four prototypes entering a lab-based user study with 32 participants addressing Research Question 2. Arrows indicate flow from prototypes to study and from study to quantitative and qualitative results.}
\end{figure}

In this paper, we adopt a mixed-methods approach (Fig. \ref{fig:studyApproach}) to design and evaluate disclosure visualizations of human–AI collaboration in journalistic news production. In part one, we ask: \textbf{(RQ1)} How can we visually represent the ratios of human-AI collaboration output in news production? To answer this, we conducted co-design sessions with designers and HCI experts (N=10) to generate disclosure design ideas. This generated 69 design concepts, exploring a wide design space of static and interactive representations. Based on usability and information visualization principles, a diverse selection was made to turn ideas into four prototypes: (a) Textual Disclosure, (b) Role-based Timeline, (c) Menu-based Chatbot, and (d) Task-based Timeline. This lead to Part 2, where we ask: \textbf{(RQ2)} How do different disclosure visualizations communicate human–AI collaboration to readers? To answer this, we ran a controlled, within-subjects lab study (N=32) combining questionnaires, eye tracking, and interviews. We examined how visualization type and human-AI collaboration dynamics influenced perceived clarity, perceived information, perceptions of AI and journalist contributions, and visual attention patterns when engaging with news articles and disclosure visualizations.

Our key findings show that (a) all prototypes effectively communicated human-AI collaboration output ratios, but for varied purposes; (b) textual disclosures were least effective in communicating human-AI collaboration; (c) Role-based and Task-based Timelines provided clearer overviews of editorial steps, while the Chatbot offered the most in-depth information; and (d) role-based timelines amplified AI contribution in primarily human-written articles, whereas task-based timeline shifted perceptions toward human involvement in primarily AI articles.

Prior research Our work offers two primary contributions: \textbf{(1)} We introduce and compare four disclosure visualization prototypes\footnote{Source code for all visualizations will be made publicly available after publication.} for representing human–AI collaboration in news production, and \textbf{(2)} We provide empirically-backed insights into how these prototypes affect reader perceptions, attention, and understanding, offering cautionary considerations on how visualization design can alter perceptions of AI’s actual role during news article creation.



\section{Related Work}
\label{sec:rel_work}

\subsection{Characterizing human-AI collaboration in journalism}
To conceptualize the relationship between humans and AI systems, prior research describes different concepts of human–AI interaction. AI assistance refers to AI providing support, emphasizing human control and responsibility \cite{zhang2020effect}. AI automation reflects varying levels of system independence in generating content with more or less editorial oversight by journalists \cite{thasler2024comes}. Other concepts focus on performance, human–AI augmentation describes AI system outperforming a human alone, whereas human–AI synergy implies the AI system outperforms either human or AI alone \cite{vaccaro2024combinations}. Moreover, human–AI co-creation is a dynamic partnership which leverages strengths towards a shared goal, emphasizing iterative, mutual learning and integration of insights \cite{kabirunleashing}, it has also been described as a strategy to amplify creativity and productivity by inspiring novel ideas \cite{JIANG2024100078interaction}. While human–AI cooperation frames AI as an independent entity that understands human preferences, working toward shared goals alongside humans \cite{li2022human,bertino2020artificial}, with a focus on aligning mental models, and communication \cite{JIANG2024100078interaction}. In the work of \citet{Breckner01072025HAIrela} cooperation is a part of collaboration, where the result does not exceed the combined individual inputs. In contrast, human-AI collaboration captures a broader, symbiotic, and adaptive relationship in which humans contribute creativity, intuition, and contextual knowledge, while AI provides computational power, data-driven insights, and efficiency, working together towards common goals \cite{saha2023human}. \citet{fragiadakis2024evaluating} extend this with a framework of tasks, goals, interaction, and task allocation. Tasks shape the collaboration type, goals can be individual or collective, communication and feedback enhance mutual understanding \cite{wang2020human}, and allocation shifts dynamically to utilize complementary strengths. \citet{holter2024deconstructing} similarly emphasize agency, interaction, and adaptation, stressing the need for flexible, balanced partnerships that foster trust, efficiency, and innovation \cite{saha2023human}. In journalism, \citet{formosa2024can} highlight how humans and AI jointly create content, focusing on continuity (how much AI-generated content is used) and responsibility (how contributions are credited). These considerations intersect with ethical concerns around trust, transparency, and authorship \cite{boni2021ethical}, particularly as increased AI involvement reduces the perception of humans as sole authors \cite{formosa2024can}. In this study, we adopt a definition of human-AI collaboration as a dynamic and adaptive partnership in which both contribute their unique strengths toward shared goals. This definition emphasizes iterative feedback, mutual learning, trust, and ethical transparency, making it especially suited to capture the multifaceted interplay of human and AI contributions in news production.  

Human–AI collaboration in journalism occurs across multiple stages in the news production workflow, from story ideation to publishing \cite{bro2016improving,shi2024generative}. At each stage, the role of GenAI varies, affecting the ratio of human versus AI contribution. For example, while creating a headline may be a limited AI contribution, writing the first draft often shift a larger share of authorship to AI. In the story ideation phase, GenAI can support idea generation or content recommendation \cite{gondwe2023exploring}. In research and information gathering, it can aggregate and analyze data or fact check \cite{shi2024generative,gondwe2023exploring}. During content creation, GenAI is used for writing drafts, generating summaries and headlines, translating text or producing images or audio \cite{shi2024generative,gondwe2023exploring}. In editing and reviewing, it can do style correction to match the news organization style and verify the content, while in publishing it can personalize feeds and customize content \cite{shi2024generative}. These tasks illustrate how collaboration ratios shift across the stages, causing the amount of contribution by GenAI in the final output to differ.

\subsection{AI disclosure representation and communication}
AI disclosure can be represented in many forms, ranging from annotations and metadata to visible labels \cite{c2paIntroducingOfficial, gamage2025synthetic}. In journalism, bylines are a traditional form of metadata used to attribute authorship, and could similarly be extended to GenAI systems \cite{shane2021deepfakes, JIA2024100093bylines}. Labeling is the most common strategy to reduce risks of GenAI, covering content with visible warnings to alert users \cite{wittenberg2024Labeling}, such as watermarks \cite{madiega2023generative}, filter tags, or platform warnings \cite{shane2021deepfakes}. Labels can reduce misinformation belief and increase trust in newsroom processes \cite{zierlabeling, burrus2024unmasking}, but their effectiveness depends on whether they pursue process-based goals (explaining how content was created) or impact-based goals (warning about potential consequences) \cite{wittenberg2024Labeling}. This research builds on process-based approaches, which communicate the human–AI collaboration ratio, while keeping a neutral stance towards the consequences of GenAI usage in news media. Showing human involvement alongside AI contributions can improve transparency and prevent assumptions that AI alone produced the content \cite{zierlabeling}. Oversimplified “AI-generated” labels decrease belief in news headlines and willingness to share them because readers assume no human input was involved \cite{altay2024people}. Additionally, effective disclosure should specify how, when, and where AI was used in the creation process \cite{burrus2024unmasking}. Readers consistently express a desire for disclosure when news is produced primarily by AI with some human oversight \cite{fletcher2024does} and want to know about AI involvement regardless of its impact on public opinion or perceptions of its accuracy, and trustworthiness \cite{piasecki2024ai}. Even though Article 50 of the EU AI Act exempts disclosure when content has been human-reviewed \cite{euaiact2024}, readers still want to know the origin of the content \cite{piasecki2024ai}. To address such shortcomings, our work aligns with the design space proposed by \citet{gamage2025synthetic}, but applies it to visually disclose the nuanced collaboration between humans and AI in journalism. We focus on iconography, and level of detail, while excluding sentiment as our approach is process-based \cite{wittenberg2024Labeling}, and positioning as our study focuses on design rather than placement.

Social media platforms showcase a range of AI disclosure strategies. Meta attaches “AI Info” labels with expandable details \cite{meta_2024}, TikTok uses “AI-generated” and “creator labeled as AI-generated” \cite{tiktok_guide}, LinkedIn employs Content Credentials metadata \cite{linkedin_2024}, and YouTube applies broader warnings such as “Altered or synthetic content” \cite{youtube_guide}. Furthermore, prior work on AI disclosure representations have had various approaches, focusing on design, context, and impact on user understanding, trust, and engagement. Label wording strongly influence perception, terms such as “AI-generated,” “Generated with an AI tool,” or “AI Manipulated” are consistently associated with AI involvement \cite{epstein_fang_arechar_rand_2023}. Users prefer pictograms and explicit details about AI usage \cite{sivakumaran2023investigating}. Alternative formulations have been tested, including “Content Generated by AI” \cite{li2024impact}, “Written by artificial intelligence" (translated from Dutch) \cite{piasecki2024ai}, and variants such as “AI-Generated” or “Altered” \cite{wittenberg2024longlabeling}. More elaborate formats include “AI Usage Cards” for standardized reporting in scientific research \cite{wahle2023ai} and certification labels that improve trust in both low- and high-stakes contexts \cite{scharowski2023certification}. Additionally, progressive disclosure can balance transparency and cognitive load by showing basic AI involvement upfront and revealing more detail upon user request \cite{springer2020progressive}. The cognitive load of the user should be taken into account, showing which elements involved GenAI, but too much technical details will be overwhelming \cite{burrus2024unmasking}. Therefore, an effective label should enable viewers to recognize the human-AI collaboration at first glance.

\subsection{Information visualization techniques for AI disclosures}\label{lit:infovis}
Information visualization is the interactive, computer-generated graphical representation of data, traditionally used to help humans understand and analyze complex datasets \cite{chen2010information, moere2011role}. Its goal is to transform information into intuitive and meaningful representations that provide insight \cite{chen2010information}. Design is inevitable to this process: visualizations must balance utility, reliability, and attractiveness, while accounting for perceptual and cognitive constraints that shape user interpretation \cite{moere2011role}. For disclosing human–AI collaboration, the main challenge is how to represent ratios clearly and effectively. Proportional visualization is common in media \cite{siirtola2014bars}, using familiar techniques such as pie and donut charts, stacked bar charts, tree maps, waffle charts, and numerical or textual formats \cite{Bianconi2024}. Pie and donut charts are widely used despite difficulties in judging areas \cite{spence1991displaying}, together with the stacked bar chart, the most common visualization techniques in media \cite{siirtola2014bars}. Tree maps extend to hierarchical data, waffle charts represent proportions as grids, and textual notations such as percentages (“75\% AI, 25\% Human”) are representations for displaying proportions \cite{cui2019text}. Although these techniques are static and simple, they provide a useful starting point. Embellishments, though sometimes seen as distracting, can in fact attract attention and support long-term memorization \cite{cui2019text}. 

Design principles from HCI, such as Nielsen’s heuristics, are prominently used in interaction design. We focus on three of these heuristics for disclosure representations: \textit{aesthetic and minimalist design}, \textit{recognition rather than recall}, and \textit{match between system and the real world} \cite{nielsen_ten_2005}. These map onto three core principles for disclosure design: simplicity, cognitive load management, and user-centered approach. These principles will ensure that users can accurately interpret AI involvement, avoid misperception, and interact with information at different levels of detail. AI disclosure visualizations are primarily informative rather than interactive, which makes many of Nielsen’s other heuristics less relevant, as they address complex user interactions and error recovery rather than the goal of creating clear, self-explanatory representations. Creating AI disclosure representations that are clear and free from unnecessary information is crucial for usability, aligning with Nielsen’s heuristic of aesthetic and minimalist design \cite{nielsen_ten_2005} and the design principle of simplicity in interaction \cite{rees2001designing}. Clarity is essential \cite{valverde2011principles}, since misleading or ambiguous representations can distort perceptions of AI’s role. When human and AI contributions are not clearly distinguished, users may overestimate or underestimate AI's role \cite{altay2024people}. Managing cognitive load is equally important, as users should immediately understand information without relying on memory \cite{nielsen_ten_2005}. Progressive disclosure provides one solution, offering detail on demand \cite{engelbrecht2015designing} which helps avoiding overload. Representations must also match the real world \cite{nielsen_ten_2005}, using familiar formats. Additionally, interactive visualizations should not only be functional, but also intuitive and engaging \cite{hurterinformation}. These heuristics align with the specific goal of creating human-AI disclosure visualizations in journalism, ensuring that the representations we develop emphasize clarity and user friendliness while taking cognitive load into account.

\section{Part 1: Designing human-AI collaboration representations}
\subsection{Co-design sessions}
Co-design sessions were conducted to generate design ideas for human-AI collaboration representations in news media. These sessions leverage the collective creativity of participants with different ideas and knowledge to come up with a design solution together \cite{sanders2008co, steen2013co}. The aim was to explore diverse static and interactive representations, where static designs present information at once, while interactive designs progressively reveal information \cite{springer2020progressive}. Participants received a sensitizing booklet beforehand, followed by the co-design session.

\subsubsection{Procedure}
\begin{figure}
    \centering
    \includegraphics[width=1\linewidth]{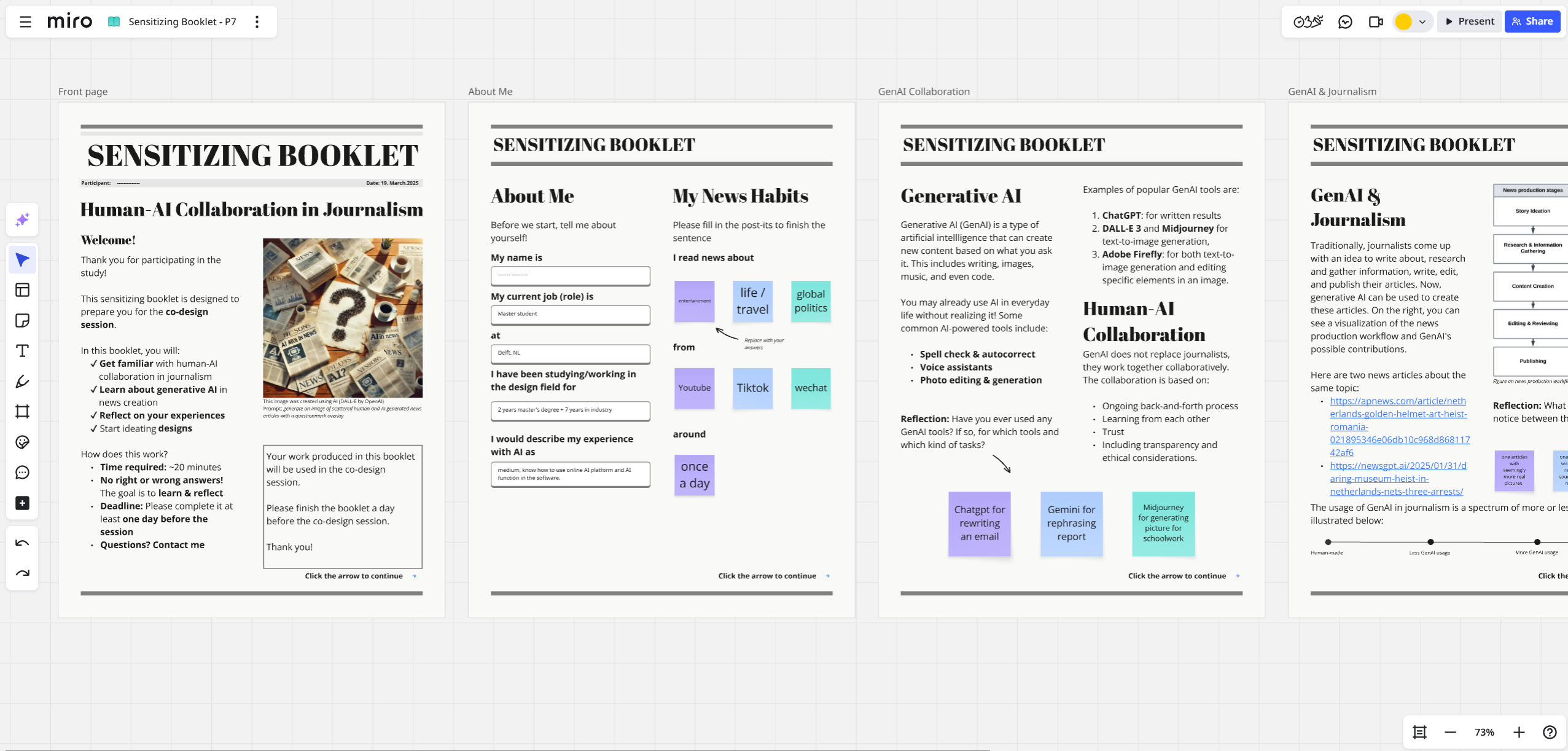}
    \caption{Screenshot of a filled out sensitizing booklet in Miro, as part of our preparation for the co-design session.}
    \label{fig:sensitizingbooklet}
    \Description{Screenshot of the Miro board showing a filled out sensitizing booklet. First page shows an introduction giving instructions and an explanation about the research. Second page shows questions and answers about the participant and news reading habits. Third page explains GenAI and human-AI collaboration with a reflection question. Fourth page explains GenAI and journalism.}
\end{figure}

\textbf{Sensitizing Booklet}. Each participant completed a sensitizing booklet ($\sim$20 min) in Miro\footnote{Anonymized Miro board of sensitizing booklet: \url{https://miro.com/app/board/uXjVLhVnry8=/?share_link_id=270962526928}} (Fig.~\ref{fig:sensitizingbooklet}) which introduced concepts such as GenAI in journalism, human–AI collaboration, and AI disclosures. While also reflecting on their experiences and increasing confidence to contribute during the sessions \cite{trischler2019co}. This was completed before the co-design session, and was submitted together with a signed informed consent form.

\begin{figure}
    \centering
    \includegraphics[width=0.6\linewidth]{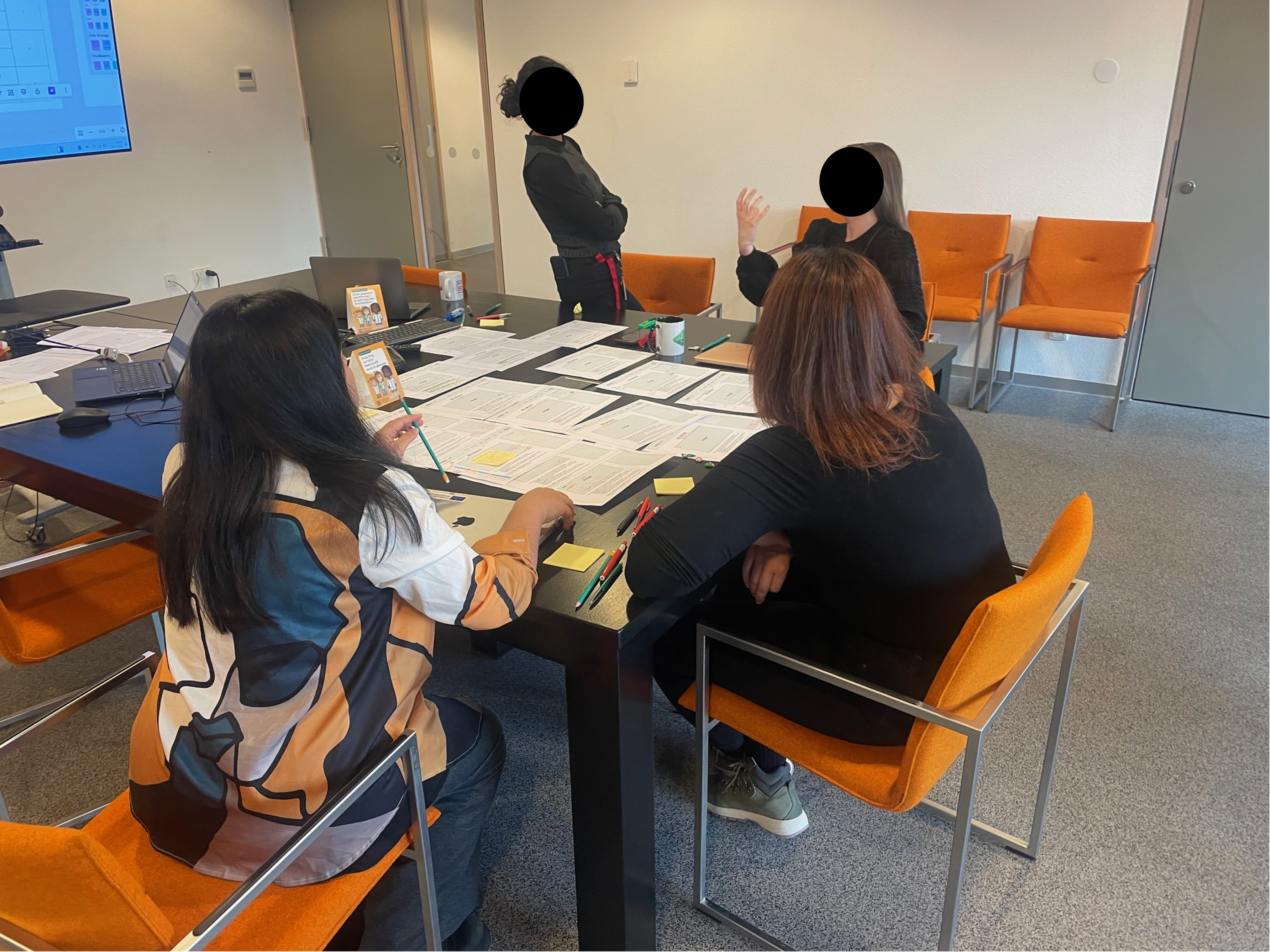}
    \caption{In-person co-design session setup.}
    \label{fig:codesignSetup}
    \Description{A picture of the in-person co-design session. It shows the designers working in teams and the templates spread out on the table with post-its, pens and pencils.}
\end{figure}

\begin{figure}
    \centering
    \includegraphics[width=0.6\linewidth]{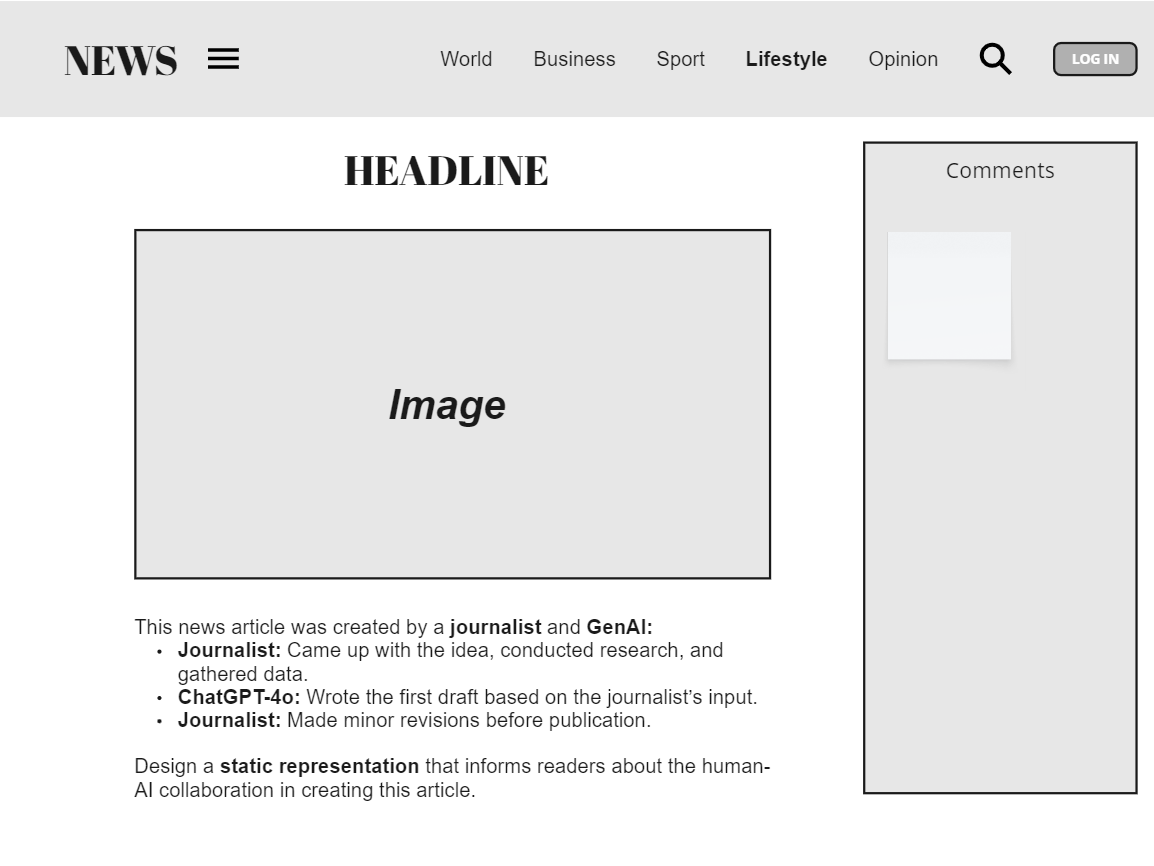}
    \caption{Example ideation template for co-design session}
    \label{fig:example-scenario1}
    \Description{Template of a news article layout used in the co-design session. The page shows a headline, a large placeholder image, and a comment section on the right. Below the image is the scenario text describing journalist and AI roles: the journalist generated the idea, conducted research, and revised the article; ChatGPT-4o produced the first draft. The template asks participants to design a static or interactive representation that communicates human–AI collaboration in creating the article.}
\end{figure}

\textbf{Co-design Session}. Three co-design sessions were then conducted (90 min each) in March 2025: two online via Zoom and Miro\footnote{Anonymized Miro board of co-design session: \url{https://miro.com/app/board/uXjVILuPkmM=/?share_link_id=330092544705}} with drawing tablets, and one in-person at XYZ institute using drawing materials (Fig.~\ref{fig:codesignSetup}). In all sessions, designers worked in pairs, combining more experienced with less experienced participants. The sessions started with an explanation of the concepts and reflection on the booklet, followed by two ideation rounds using scenarios of varying AI contributions in the journalistic workflow, in between these rounds they were shown requirements and Information Visualization techniques to inspire, but not restrict them. During the ideation rounds they received ideation templates (Fig.~\ref{fig:example-scenario1}) which looked like a wireframe of a news website with a scenario, after these rounds there was a short feedback round to foster collaboration between the teams. 

\subsubsection{Participants}
Ten designers were recruited through snowball sampling. Three were (UX) design researchers, two were HCI experts, and five were design master's students. The participants that joined the online session were required to have access to a drawing tablet. To focus on communication design rather than newsroom workflow, journalists were deliberately excluded. Each participant joined a single session and was compensated with a \$/€20 gift card.

\subsubsection{Analysis and findings}
\begin{figure}
        \includegraphics[width=0.49\textwidth]{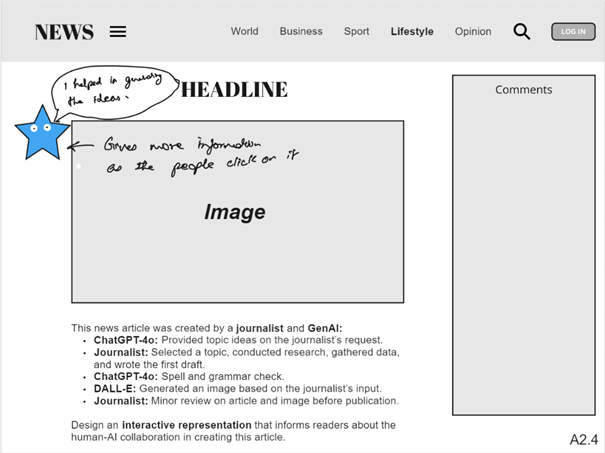}
    \hfill
        \includegraphics[width=0.49\textwidth]{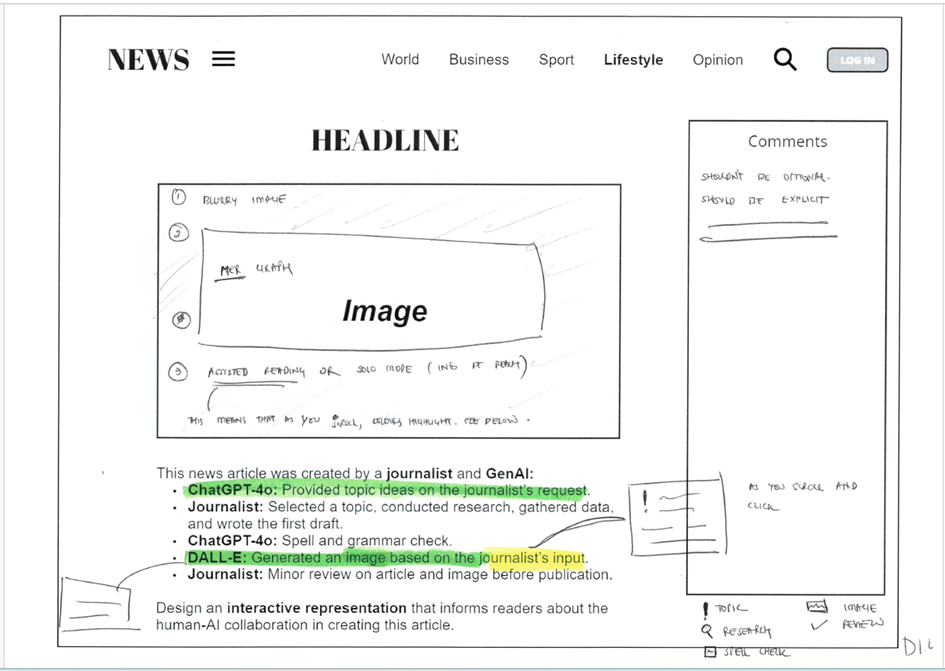}
\caption{Two examples of the results of the co-design sessions, drawn on the ideation template. The example on the left was created online in Miro, the example on the right was digitized from the in-person session.}
\label{fig:codesignResult}
\Description{Two examples of the output produced by designers during the co-design sessions. Both show a wireframe of a news website with a headline, image placeholder, a scenario text, and a comment section on the right. The design on the left shows a star icon which can be clicked, and will then display information. The right template contains two designs: a blurred image that needs to be clicked to be revealed, and highlighted text which shows information when hovered.}
\end{figure}

All sketches (digital and scanned) (Fig.~\ref{fig:codesignResult}) were collected, transcribed into descriptions, and inductively coded \cite{braun2012thematic}. Duplicate ideas were merged, resulting in 69 unique concepts: 29 static and 40 interactive. Static designs fell into three categories: (1) labels/icons/stamps, (2) textual explanations, and (3) more elaborate visualizations such as bar or gauge-like indicators. Interactive designs were grouped into six categories: (1) sliders/adjustable, (2) click/toggle, (3) hover/reveal, (4) pop-ups, (5) scroll/floating elements, and (6) chatbot-style disclosures. Placement varied widely, with ideas situated near headlines, images, article bodies, or underneath the article. A full overview of the results can be found in Supplementary Materials A. These ideas provided a wide design space for visualizing human-AI collaboration ratio's from which prototypes for evaluation were selected.

\subsection{Design selection}\label{sec:designeval}
From the 69 design ideas generated, we selected four prototypes to use in the controlled user study. The set was purposefully limited to four prototypes to balance the need for diverse designs with the practical constraints of a lab-based user study, including experiment duration, multiple measurements, and participant attention. The aim was to capture diversity in modality (visual vs. textual) and interactivity (static vs. interactive), building on the design space proposed by \citet{gamage2025synthetic}. Modality and interactivity map onto usability principles of using familiar formats and progressive disclosure, ensuring clarity, feasibility, and communicative value (Sec.~\ref{lit:infovis}).

Selection criteria (Table \ref{tab:designcriteria}) were grounded in HCI and Information Visualization principles \cite{quispel2018aesthetics, nielsen_ten_2005} and practical constraints of lab-based testing. Designs had to represent both human and AI contributions, be technically and journalistically feasible, minimize cognitive load, and avoid unnecessary complexity. This process was conducted collaboratively by two authors.

Based on these criteria, four designs were selected: \textbf{(1) Textual Disclosure (TD)} – a label-like textual explanation of GenAI model and its contributions, \textbf{(2) Role-based Timeline (RT)} – icons of person and robot to show division of work, \textbf{(3) Chatbot (C)} – a menu-based chatbot with predefined questions, and \textbf{(4) Task-based Timeline (TT)} – a workflow timeline with interactive hover element. These four design ideas were turned into working prototypes for user testing. 

\begin{table}
\centering
\small
\renewcommand{\arraystretch}{1.3}
\begin{tabularx}{\textwidth}{lX}
\hline
\textbf{Criterion} & \textbf{Description} \\
\hline
Human–AI collaboration & Must represents both human and AI contributions, not only AI usage. \\
Feasibility (technical) & Can be implemented as a working prototype for the study. \\
Feasibility (journalistic) & Realistic to integrate into journalistic workflow. \\
Simplicity & Clear \cite{quispel2018aesthetics}, free from unnecessary elements, and immediately understandable \cite{nielsen_ten_2005}. \\
Cognitive load management & Users should be able to grasp the AI-human ratio without needing to remember prior  information or interpret complex details \cite{nielsen_ten_2005}. \\
User-centered approach & The visualization should align with user expectations and avoid unnecessary technical complexity \cite{nielsen_ten_2005}. \\
Low interaction count & Minimizes number of required interactions, which aligns with simplicity and reduces cognitive load. \\
\hline
\end{tabularx}
\caption{Criteria used for evaluation and selection of design ideas for prototyping.}
\label{tab:designcriteria}
\end{table}

\subsection{Prototype development}\label{sec:protodev}
The four selected design ideas were turned into fully functional prototypes (Fig.~\ref{fig:teaser}). Each was developed in JavaScript, JSON, and CSS for integration in a React-based web application. All visualizations are similar in size for consistency and easy comparison. 

The \textbf{TD} describes the used AI model, its contributions, and the journalist's contributions. The \textbf{RT} visualization shows a linear progression of the roles between the journalist and AI, including icons as visual elements with a textual explanation underneath. The AI is displayed as a robot, this was also often done by the designers in the design sessions. The \textbf{C} disclosure is menu-based with predefined questions about the AI model, and human and AI contributions. The \textbf{TT} shows the five stages of news article creation, visualized with icons. Human-AI collaboration is indicated with a sparkle symbol, a neutral label commonly seen in AI context \cite{gamage2025synthetic}. When users hover over the icons, additional information about the contributions is revealed. Each visualization went through multiple design iterations. The eight developed prototypes (4 disclosure visualizations x 2 collaboration ratios) are included in Supplementary Material~A, the four disclosure visualizations with high AI contributions are displayed in Fig.~\ref{fig:teaser}.

\section{Part 2: Evaluating human-AI collaboration disclosure visualizations}
To assess the effectiveness of the selected disclosure designs, we conducted a controlled lab-based user study. The aim was to investigate how users perceive, interpret, and interact with the different disclosure visualizations of human-AI collaboration in online news articles. A mixed methods study was conducted, combining quantitative data from eye tracking, and questionnaires, with qualitative data from semi-structured interviews. 

Eye tracking was included to understand how users visually engage with the disclosure visualizations. The self-reported questionnaires capture how participants perceived the disclosures and semi-structured interviews reveal deeper insights into their understanding, while gaze measures offer complementary insights on attention and information processing \cite{holmqvist2012eye, albert2022measuring, andrychowicz2018basic}. This mixed methods approach allows us to explore whether participants understood the disclosures and how they visually navigate them, offering a comprehensive understanding on the communicative effectiveness in the news context.

\subsection{Study design}
Our study follows a 4 (IV1: disclosure visualizations: TD, RT, C, TT) x 2 (IV2: human-AI collaboration ratio: Primarily Human vs. Primarily AI) within-subject study design.  The trials were created using two news articles. For each article, a primarily human and primarily AI contribution version was created (1-Human, 1-AI, 2-Human, 2-AI). Each of these four article versions was paired with each of the four disclosure visualizations (TD, RT, C, TT), so every disclosure visualization appeared with every article version. Therefore, each participant completed 16 trials (4 article versions x 4 visualizations), with disclosure visualizations and article version order counterbalanced using a 4x4 Latin square to minimize order effects. All sessions were conducted between May and June, 2025 at XYZ institute. The sessions took approximately one hour, and no tasks prior to the experiment were asked of the participants. 

\subsubsection{Measures}
In this study, quantitative measures consisted of the following dependent variables: 
(a) Perceived human–AI collaboration, adapted from IBM’s AI Attribution Toolkit \cite{AIAttributionToolkitIBM}, capturing the communicated proportion of human vs. AI contributions.
(b) Perceived AI role per task, based on Altay et al. \cite{altay2024people}, using 7-point Likert ratings for AI involvement in tasks such as topic selection, writing, editing, fact-checking, and headline creation.
(c) Perceived clarity and value, 7-point Likert-scale adapted from Wittenberg et al. \cite{wittenberg2024longlabeling}, including items such as clear, helpful, easy to understand, useful, and informative.
(d) Perceived information, 7-point Likert ratings of whether the visualization conveyed overview, in-depth information, specific workflow steps, or who contributed what.
(e) Trial duration, logged automatically from the moment a disclosure appeared until the questionnaire was submitted. Although differences were expected between static and interactive visualizations, this measure provided insight into the time needed for each disclosure type, which may influence real-world adoption.
(f) Eye-tracking metrics, recorded within predefined Areas of Interest (AOIs): total gaze duration, mean fixation duration, fixation count, saccade count, and mean saccade length \cite{holmqvist2012eye}. 

These measures assess whether readers understood who contributed what to the article, such understanding improves transparency, can reduce misinformation belief and increase trust in newsroom processes \cite{zierlabeling, burrus2024unmasking}. If a disclosure fails to communicate human-AI collaboration clearly, it cannot fulfill its transparency purpose, and its effect on authorship \cite{hwang202480}. Combined with measures of clarity, information type, and visual engagement of the disclosures, they create a comprehensive assessment of disclosure effectiveness. Therefore, these measures are not manipulation checks but core outcomes that indicate whether a disclosure visualization communicates the information needed for meaningful transparency in journalism.

Qualitative data were gathered after the trials. A consensually audio-recorded semi-structured interview was conducted, focusing on their overall interpretation of disclosures, design preferences, and real-world application. The full list of interview questions can be found in the Supplementary Material~B. Our study followed strict guidelines from our institute’s ethics and data protection committee.

\subsection{Study procedure}
\begin{figure}
    \centering
    \includegraphics[width=1\linewidth]{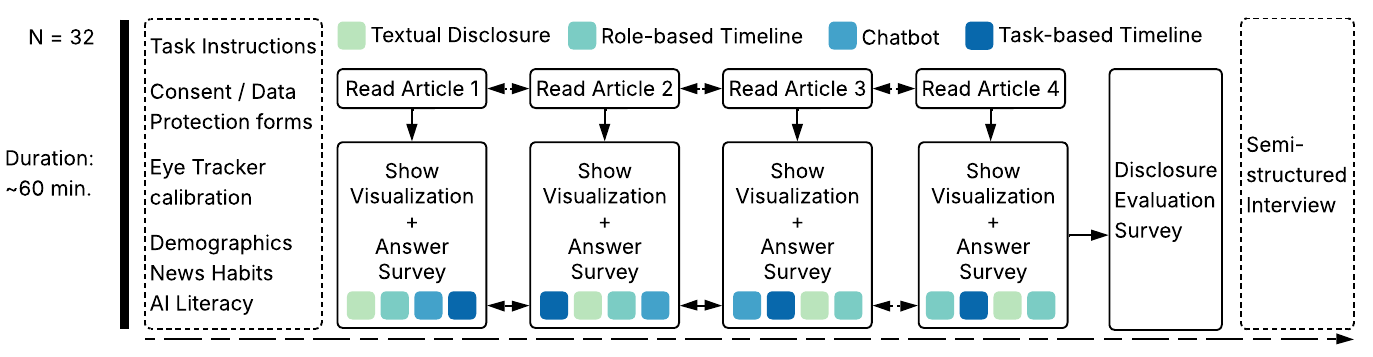}
    \caption{User study procedure. The four Articles refer to the four article versions (1-Human, 1-AI, 2-Human, 2-AI), counterbalanced between participants using a Latin square.}
    \label{fig:UserStudyApproach}
    \Description{Flow diagram of the user study procedure with 32 participants lasting approximately 60 minutes. The session began with task instructions, consent forms, eye tracker calibration, and questionnaires on demographics, news habits, and AI literacy. Participants then read four articles, each followed by the four disclosure visualizations, and a survey. The order of visualizations was rotated across articles and participants. After completing all sixteen trials, participants filled out a disclosure evaluation survey and ended with a semi-structured interview.}
\end{figure}

Our study procedure is shown in Fig.~\ref{fig:UserStudyApproach}, and lasted approximately 60 minutes. Participants first read and signed the informed consent form, and filled in a short pre-task questionnaire about demographics, news consumption habits, and their AI experience. Vision screening questions ensured participants did not have impairments that could interfere with eye tracking.

A verbal overview of the study was provided, explaining the structure of the trials and how to navigate the web interface. Participants were informed that they would be reading news articles and evaluating disclosure visualizations, but were not told the specific disclosure visualizations in advance to maintain a natural first impressions. The Tobii Pro Fusion eye tracker was calibrated at the start of each session.

Following calibration, the main experiment task started, consisting of 16 trials (4 article versions × 4 visualizations). Each trial followed the same order: (1) the news article appeared without disclosure, (2) the disclosure visualization was displayed beneath the article, and (3) a questionnaire appeared on the right side of the screen. Participants continued through these steps by pressing a button on the interface. This sequential order ensured clear mapping of eye tracking data to pre-defined Areas of Interest (AOIs) \cite{hessels2016area}.

After completing the 16 trials, participants filled out a short questionnaire evaluating the four disclosure visualizations on qualities such as perceived clarity, value and informativeness. They then took part in a semi-structured interview, where they reflected on their preferences, interpretation of the disclosures, and possible use in real-world news contexts. All interviews were audio-recorded for later transcription and thematic analysis. Participants were compensated with a \$/€10 gift card.

\subsubsection{Participants}
32 participants\footnote{For effect size $f$=.25 under $\alpha$=.05 and power (1-$\beta$)=.95, with 16 repeated measurements within factors, one would need a minimum of 15 participants.} (19 female, 12 male, 1 non-binary) were recruited through multiple platforms and snowball sampling. Ages ranged from 22 to 72. Educational backgrounds were varied (completed high school, bachelor, master, and doctorate), and news consumption habits ranged from daily to rarely. AI literacy, measured by an adapted\footnote{We changed the term in Q5 from `devices` to `technology` to ensure the scale is meaningful to participants, since we do not focus on devices per se.} subset of items from the modular 11-point MAILS scale \cite{carolus2023mails}, averaged 5.69 (SD = 1.58). The MAILS items were selected from four dimensions: AI understanding, detection, ethical awareness, and persuasion literacy. These align most closely to how readers interpret and evaluate AI involvement in news production (the questionnaires can be found in Supplementary Material~B). Cronbach’s $\alpha$=.87 showed good internal consistency for the selected items \cite{tavakol2011cronbach}. Average self-reported ChatGPT experience on a 1–7 scale (7 = very experienced) was 4.22 (SD = 1.96). The only requirement was that participants did not have significant visual impairments.

\subsection{Hardware and software setup}
\subsubsection{Hardware} Participants were seated at a desk with a 25-inch monitor at a resolution of 2560 x 1440 pixels, with a Tobii Pro Fusion screen-based eye tracker\footnote{\url{https://www.tobii.com/products/eye-trackers/screen-based/tobii-pro-fusion}} mounted below the screen (Fig.~\ref{fig:UserStudySetup}). The system was connected to a laptop running Tobii Pro Lab, which was used for calibration, data collection, and later analysis.

\subsubsection{Software} 
\begin{figure}
        \includegraphics[width=0.40\textwidth]{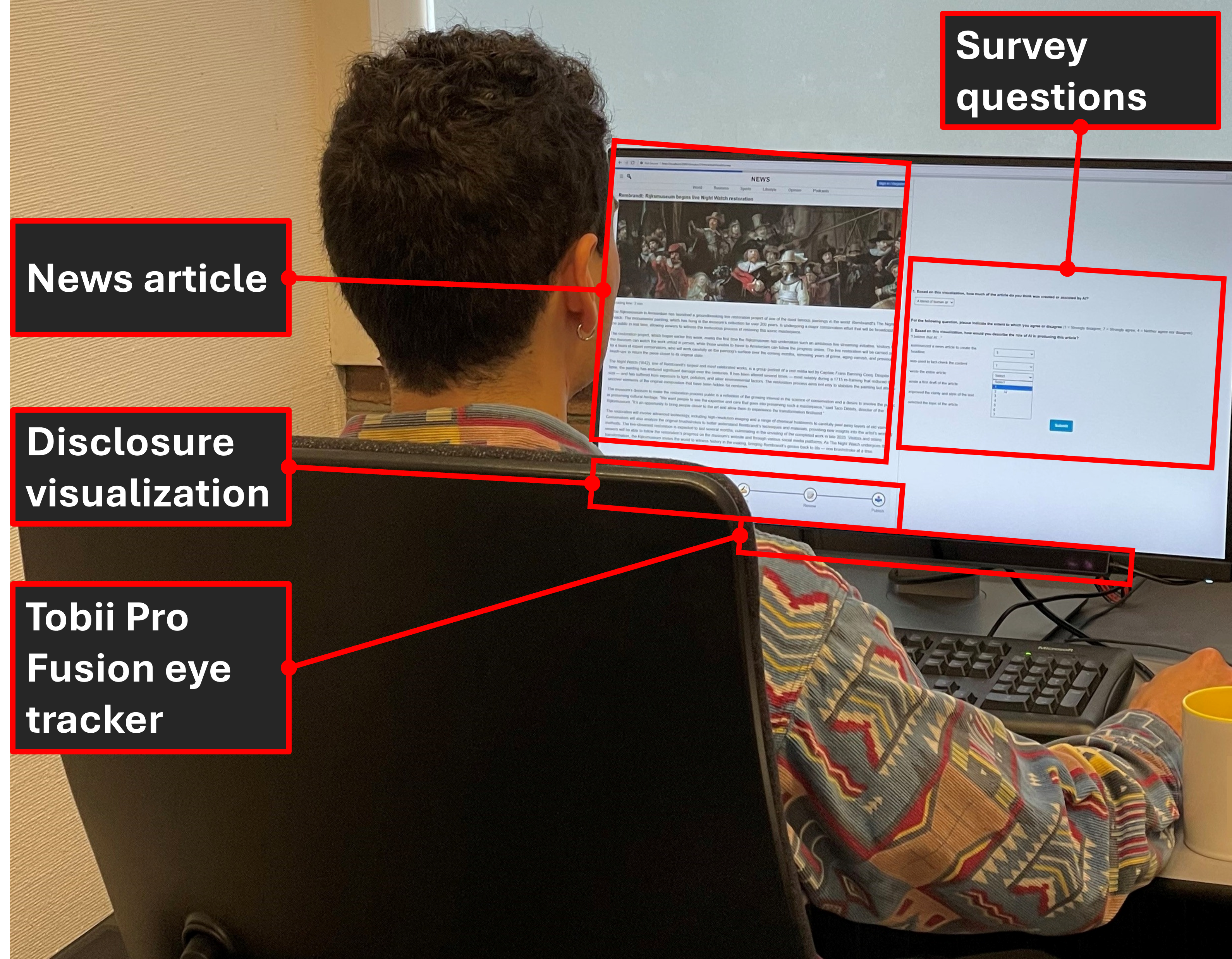}
    \hfill
        \includegraphics[width=0.59\textwidth]{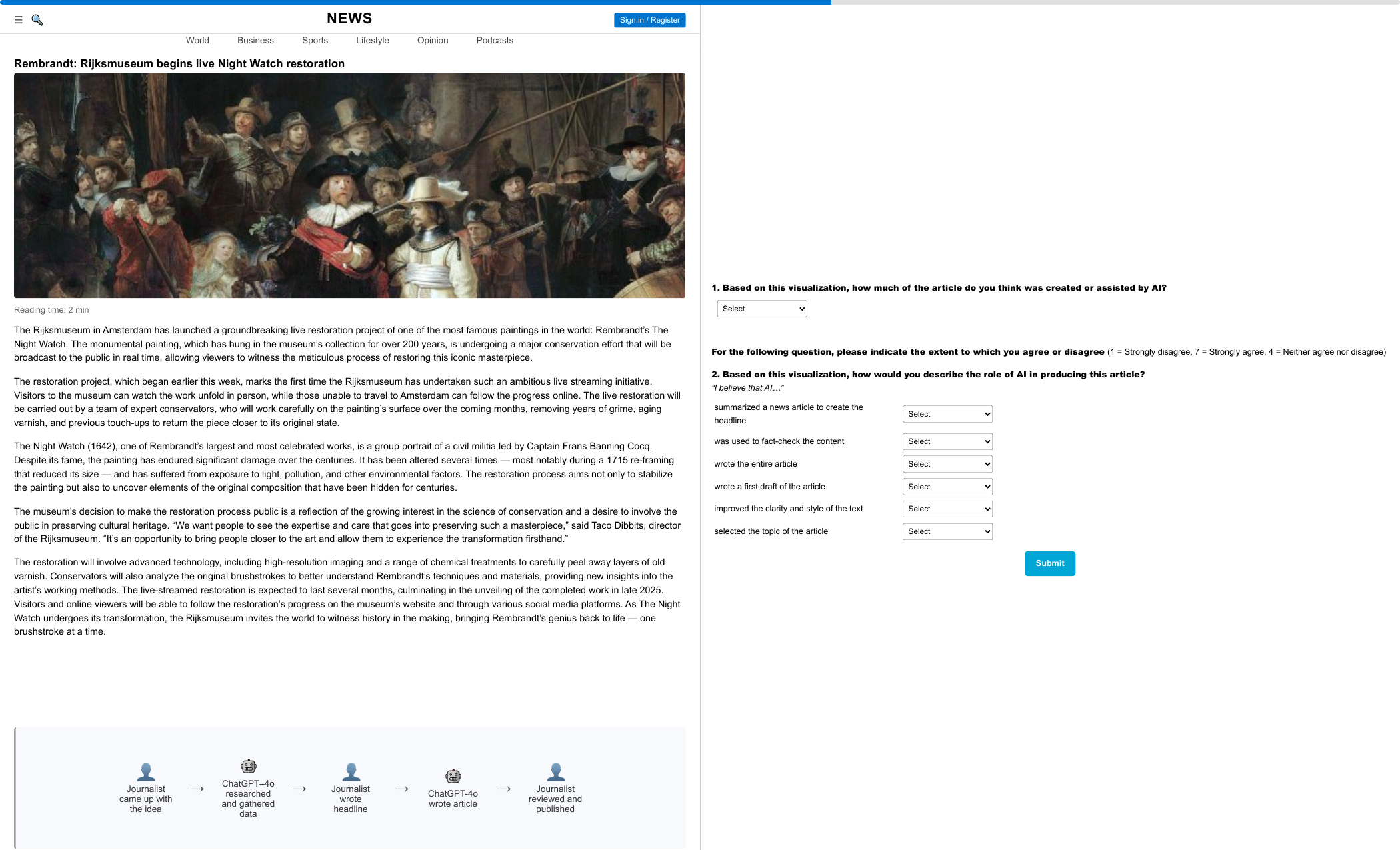}
\caption{User study setup}
\label{fig:UserStudySetup}
\Description{On the left side, the participant is shown from behind with the screen displaying a news article, disclosure visualization, and survey. The Tobii Pro Fusion eye tracker is attached underneath the monitor. On the right side, a close-up of the interface shows the article text with an image at the top, disclosure visualization below, and survey questions to the right.}
\end{figure}

Stimuli were presented via a custom built React web application hosted locally, to present the stimuli and collect the questionnaire responses. The interface divided the screen into two panes: the left displayed the stimulus (article and visualization) in a realistic news website format, and the right contained the questionnaire (Fig.~\ref{fig:UserStudySetup}).

Two real low-stake news articles were selected from the MiRAGeNews dataset \cite{huang2024miragenews}, originally published by BBC\footnote{\url{https://www.bbc.com/news/world-europe-48905867}, and \url{https://www.bbc.com/news/world-49760240}}. Both articles were adapted into two versions to manipulate the human-AI collaboration ratio, adapted from the method of Gilardi et al. \cite{gilardi2024willingness}: 
\begin{enumerate}
    \item Primarily human-contributions (Article 1-Human and Article 2-Human): The original article was used as input for ChatGPT-4o to generate a headline and change the style. This prompt mimicked the Gilardi et al. \cite{gilardi2024willingness} prompt style: “\textit{Here is a newspaper article: [article text]. As an AP journalist, write a headline for this article and do small adjustments to the text to keep it in AP style.}” This generated an article that was almost entirely written by the journalist, but still collaborating with GenAI.
    \item Primarily AI-contributions (Article 1-AI and Article 2-AI): Only the article headline was provided to ChatGPT-4o, using an adapted prompt from Gilardi et al. \cite{gilardi2024willingness}: “\textit{As an AP journalist, you must write a 400-word article on this topic: [article headline].}” This generated an article that was almost entirely written by GenAI.
\end{enumerate}
The full news articles used as stimuli can be found Supplementary Materials~B. Each version was paired with the four prototypes (TD, RT, C, TT; see Fig.~\ref{fig:teaser}), creating 16 unique stimuli. The information represented in these prototypes were derived from these controlled prompt-based manipulations to create the article versions. Disclosure visualizations were consistently placed at the bottom of the article to control for placement effects.


\subsection{Quantitative results}
We report our analysis of time spent on task, perceived human-AI collaboration, AI role, perceived clarity and value, perceived information, and eye tracking metrics. The ordinal response variables (perceived human-AI collaboration, AI role, and clarity and informativeness ratings) were analyzed using cumulative link mixed-effects models (CLMMs) \cite{christensen2018cumulative} to take the ordered nature of a Likert-scale into account. Continuous variables (time spent and eye tracking metrics) were analyzed using linear mixed-effects models (LMM) \cite{BatesLMM}. Both models included participants as random intercepts to model the repeated measures design \cite{Kaptein2016}, and self-reported AI literacy (scale 0-10) and ChatGPT experience (scale 1-7) as continuous covariates. Interactivity (static vs. dynamic visualization) was not included as a random effect, as two-level grouping factors do not provide enough information to reliably estimate random-effect variance \cite{OberprillerRandomEffects}. These mixed-effects models allowed us to model individual differences while testing fixed effects of visualization type and collaboration condition where applicable. To control for potential inflation of Type I Error, False Discovery Rate (FDR) correction \cite{haynes2013benjaminiFDR} was applied to adjust the $p$-value across the six perceived AI role features, the ten perceived clarity, value, and information features, and for all post hoc comparisons. Due to the exploratory nature of this study, FDR was chosen instead of more conservative Bonferroni or Tukey adjustments. Effect sizes for CLMMs are reported as odds ratios (OR) with 95\% confidence intervals (CI95\%) for pairwise contrasts. For LMMs, effect sizes are reported as partial $\eta^2$ for main effects and Cohen’s $d$ for pairwise contrasts \cite{cohen2013statistical}. 

\begin{figure}
    \includegraphics[width=\linewidth]{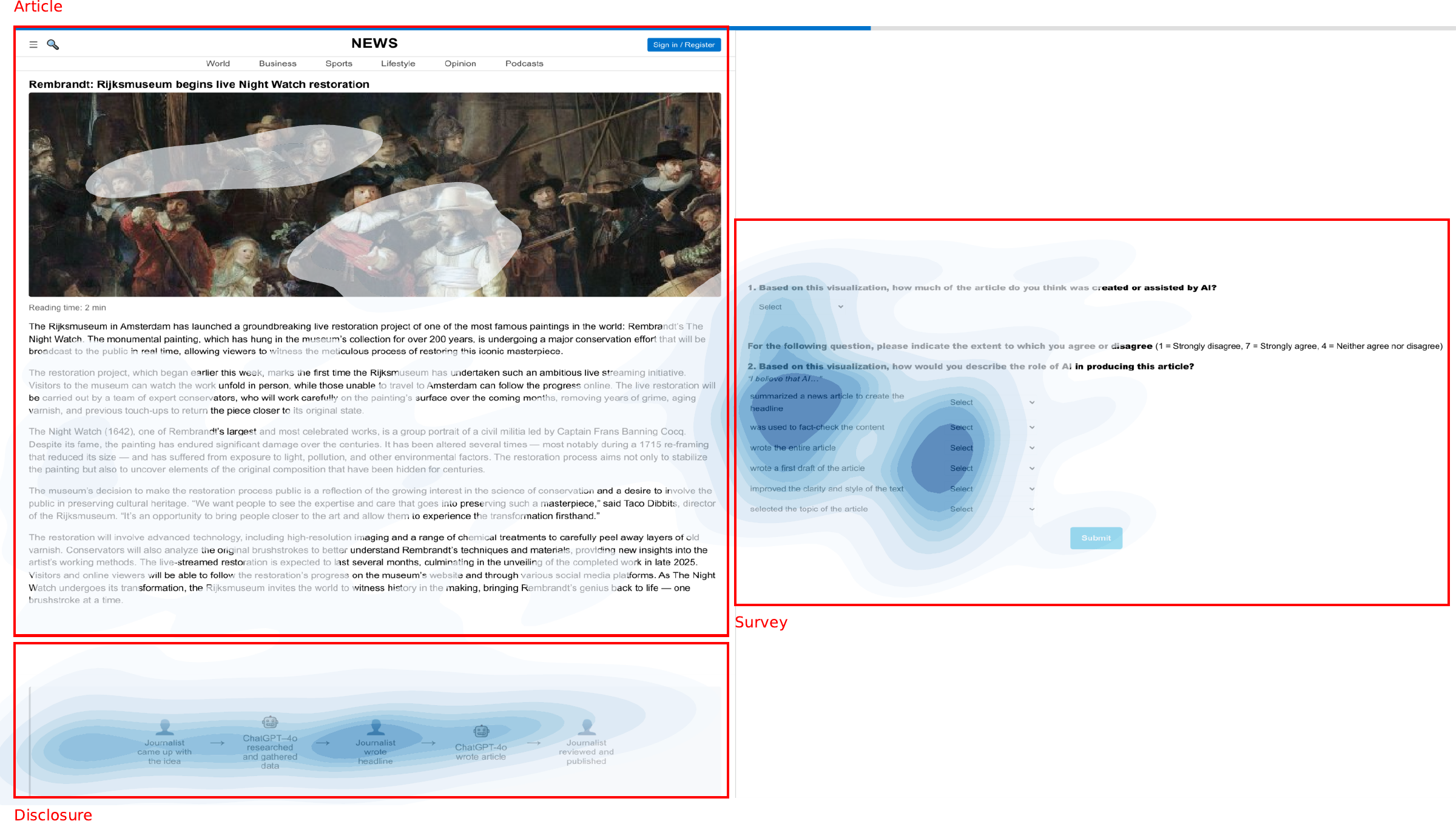}
    \caption{Heatmap of the gaze points of one participant on a stimulus. The red boxes indicate the predefined AOIs.}
    \label{fig:gaze_heatmap}
    \Description{Stimulus screen with an overlaid gaze heatmap from one participant. The layout shows a news article with image and text on the left, a disclosure visualization below, and survey questions on the right. Red boxes outline predefined Areas of Interest: article, disclosure, and survey. The heatmap indicates gaze on the article text, parts of the image, disclosure visualization and the survey area.}
\end{figure}

Eye tracking data were collected using a Tobii Pro Fusion with Tobii Pro Lab, and only included when disclosure visualizations were visible, during both initial display and questionnaire answering. Only raw data inside the disclosure AOI (Fig.~\ref{fig:gaze_heatmap}) were processed with the Tobii I-VT (Fixation) filter, which classifies gaze movements slower than 30$^\circ$/s as fixations and faster as saccades, with additional parameters to remove short fixations and merge adjacent ones \cite{olsen2012tobii}. Data were considered valid if at least 80\% of gaze points were valid \cite{tobii2011accuracy}, data points were included if a valid measure was available from at least one eye \cite{hooge2019gaze}. Validity was assessed per participant, leading to the exclusion of eight. Tobii defines good accuracy as below 0.8$^\circ$ and precision SD below 1.5$^\circ$ \cite{tobii2011accuracy}, but due to the large Areas of Interest we adjusted the average accuracy threshold to 1.6$^\circ$, resulting in a final dataset of 20 participants. CLMM and LMM results for all response variables and gaze features on disclosure visualization main effect are shown in Table~\ref{tab:ARTResults}. Full contrast test tables are provided in Supplementary Material~C.

\begin{table}[ht]
\centering
\small
\begin{tabular}{lrrlr}
  
  \hline
Variable & $\chi^2$ & $df$ & $p$ \\ 
  \hline
Clarity & 17.77 & 3 & 0.003 ** &\\ 
Informativeness & 13.60 & 3 & 0.018 * &\\ 
Helpfulness & 0.91 & 3 & 0.935 & \\ 
Easiness to understand & 28.02 & 3 & <.001 *** &\\ 
Usefulness & 2.27 & 3 & 0.890 & \\ 
\hline
Overview & 33.78 & 3 & <.001 *** \\ 
In-depth information & 41.48 & 3 & <.001 *** \\ 
Understanding of specific steps & 26.43 & 3 & <.001 *** \\ 
Who contributed what & 5.03 & 3 & 0.637 \\ 
Sense of AI vs. human input & 4.25 & 3 & 0.642 \\ 
\hline
Variable & $F$ & $Df$ & $p$ & $\eta^2p$ \\ 
   \hline
   Gaze duration & 31.42 &   3 & $<$.001 *** & 0.62 \\
  Fixation duration & 45.96 & 3 & <0.001*** & 0.71 \\ 
  Fixation count & 44.85 & 3 & <0.001*** & 0.70 \\
  Saccade count & 34.44 & 3 & <0.001*** & 0.64  \\
  Saccade length & 11.11 & 3 & <0.001*** & 0.37 \\ 
   \hline
\end{tabular}
\caption{Cumulative link mixed model results for response variables and linear mixed model results for gaze features, testing main effect of disclosure visualization. P values were adjusted with FDR corrections.}
\label{tab:ARTResults}
\end{table}

\subsubsection{Mean time spent per trial}\label{sect:meantime}
We first looked at the time participants spent during each trial (i.e. from the moment the disclosure was displayed until the survey was submitted for that disclosure), as an indicator of potential disengagement or abandonment. As expected, interactive disclosures require more time than static disclosures. Therefore, the mean trial duration is reported to contextualize readers engagement rather than as a performance outcome. On average, time spent per trial across disclosure visualization conditions (in seconds) were: Chatbot (M=150, SD=126), Task-based Timeline (M=111, SD=110), Textual Disclosure (M=116, SD=123), Role-based Timeline (M=110, SD=108). A linear mixed-effects model was used to account for repeated measures and individual differences in AI literacy and ChatGPT experience. Results revealed significant main effects of disclosure visualization, $F(3,474)=3.55, p=.014$. Post hoc contrasts showed that the Chatbot trials took significantly longer compared to all other conditions (vs. TD: $p$=0.035, $d$=0.30, vs. RT: $p$=0.020, $d$=0.35, vs. TT: $p$=0.020, $d$=0.34). Additionally, both covariates had a significant influence, higher AI literacy was predicted a  longer time spent per trial ($\beta$=11.64 sec, $p$=.012), whereas more ChatGPT experience predicted less time spent ($\beta$=–9.52 sec, $p$=.011).

\subsubsection{Perceived human–AI collaboration dynamics}
\begin{figure}
    \centering
    \includegraphics[width=0.5\linewidth]{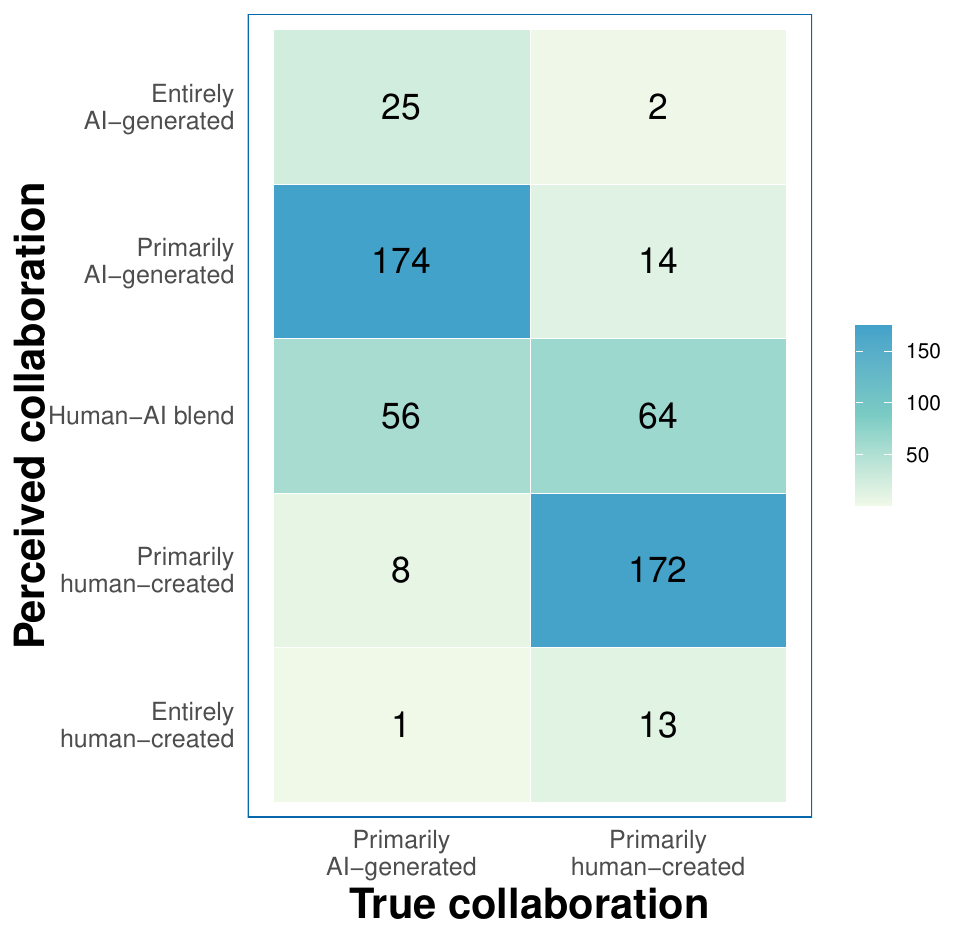}
    \caption{Confusion matrix of human-AI collaboration per article}
    \label{fig:perceived_HAICollab_confmatrix}
    \Description{Confusion matrix showing perceived versus true levels of human–AI collaboration. The X-axis represents true collaboration (primarily AI-generated, primarily human-created), and the Y-axis represents participant perceptions (from Entirely AI-generated to Entirely human-created). Cells are shaded by frequency, with darker blue indicating higher counts. With 174 cases correctly perceived as Primarily AI-generated articles and 172 cases correctly perceived as Primarily human-created articles. Misclassifications include 56-64 cases perceived as human–AI blend, and lower values for Entirely human or AI generated.}
\end{figure}

How the human-AI collaboration was perceived compared to the true collaboration ratios is shown in Fig.~\ref{fig:perceived_HAICollab_confmatrix}. As expected, primarily AI-generated articles (1-AI and 2-AI) were mostly recognized as such, and primarily human-written articles (1-Human and 2-Human) were typically seen as “Primarily human-created.” Across article versions, there were responses for “Human–AI blend,” indicating that some participants assumed creation with an even blend of human and AI contributions \cite{AIAttributionToolkitIBM}.

\begin{figure}
    \centering
    \includegraphics[width=0.7\linewidth]{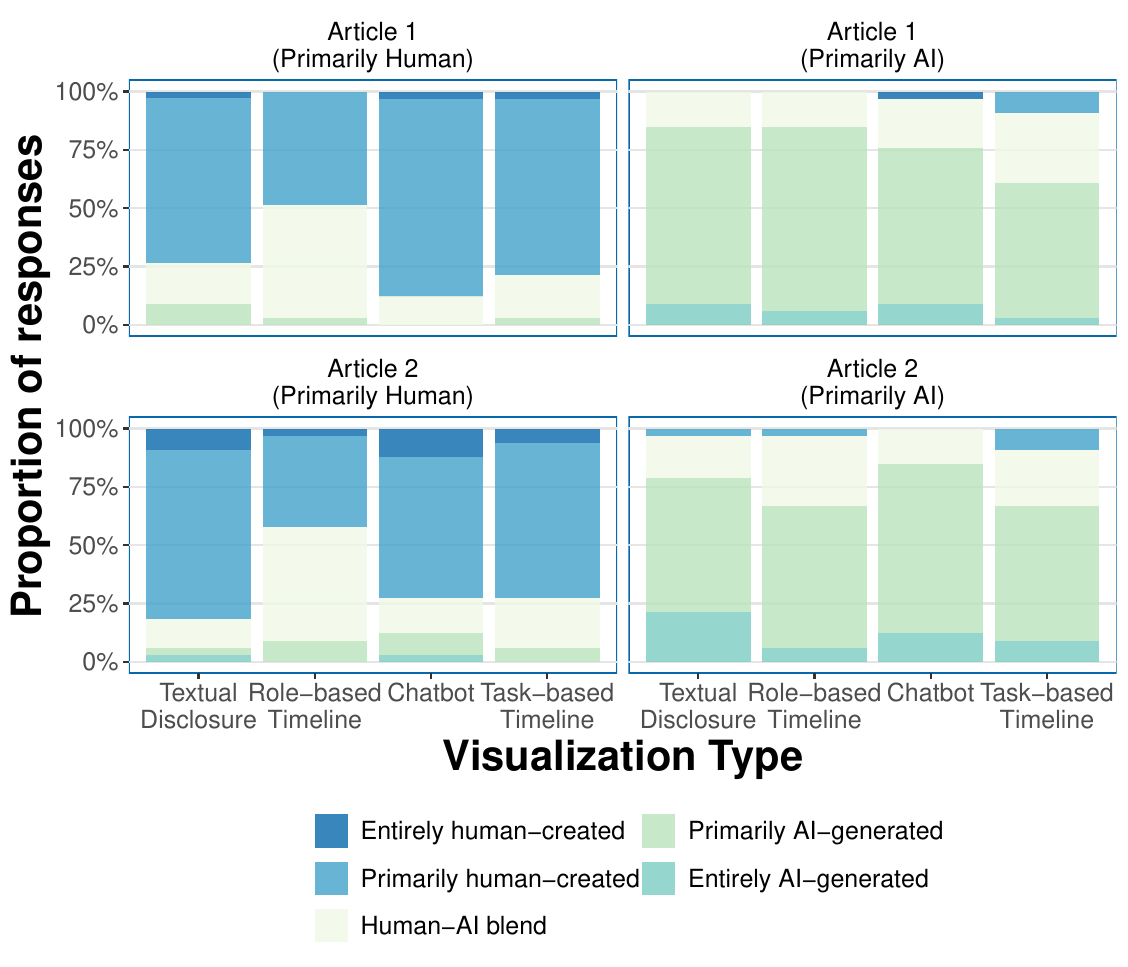}
    \caption{Proportion of responses of perceived human-AI collaboration per article and disclosure visualization}
    \label{fig:perceived_HAICollab_barplot}
    \Description{Stacked bar charts showing perceived human–AI collaboration for four articles, separated by disclosure visualization. Each chart represents one article: Articles 1 and 3 were primarily human-created, Articles 2 and 4 primarily AI-generated. The Y-axis shows proportion of responses, and the bars are divided into categories ranging from entirely human-created to entirely AI-generated. For Primarily Human articles, most responses cluster in the human-created categories, though Role-based Timeline shows more variation. For Primarily AI articles, responses were mostly on AI-generated categories, with Task-based Timelines showing more variation.}
\end{figure}

We grouped the responses into three categories (Human, Blend, AI) to ensure sufficient counts per category and analyzed with a cumulative link mixed model (CLMM) to account for the ordinal categories and repeated measures. Pairwise contrasts revealed context-dependent shifts. For primarily human articles, the Role-based Timeline significantly increased perceptions of AI contribution relative to the Textual Disclosure ($\beta=-1.30, SE=0.40, z=-3.26, p=.002, OR(95\%CI)=0.27 (0.08, 0.95))$, Chatbot ($\beta=1.62, SE=0.42, z=3.82, p<.001, OR(95\%CI)=5.06 (1.35, 19.02)$), and Task-based Timeline ($\beta=1.25, SE=0.39, z=3.19, p=.002, OR(95\%CI)=3.49 (1.02, 11.92)$). For primarily AI articles, the Task-based Timeline nudged perceptions toward human involvement, showing significant differences with Textual Disclosure ($\beta=1.05, SE=0.44, z=2.42, p=.022, OR(95\%CI)=2.86 (0.73, 11.14)$), Chatbot ($\beta=0.95, SE=0.43, z=2.21, p=.036, OR(95\%CI)=2.58 (0.68, 9.79)$), but not compared to Role-based Timeline. Fig.~\ref{fig:perceived_HAICollab_barplot} illustrates the response distributions, where these shifts appear most clearly for the Role-based Timeline (increasing Blend/AI responses in primarily human texts) and the Task-based Timeline (increasing Human responses in primarily AI texts). These findings validate that these disclosure visualizations can effectively convey nuanced impressions of human–AI collaboration, highlighting their potential to communicate underlying dynamics beyond simple labels. Yet, they may also introduce bias perceptions, shifting toward more or less AI involvement depending on the context.

\subsubsection{Perceived AI role per article}

\begin{table}
\centering
\small
\setlength{\tabcolsep}{4pt}

\begin{tabular}{l l r r l}
\hline
Response Variable & Factor & $\chi^2$ & $df$ & $p$ \\
\hline
\multirow{5}{*}{Headline}
   & \textbf{Visualization} & 4.03 & 3 & 0.259 \\ 
   & \textbf{Collaboration} & 256.56 & 1 & < .001 *** \\ 
   & \textbf{Visualization × Collaboration} & 5.97 & 3 & 0.141 \\ 
   & \textbf{\textit{AI literacy}} & 9.32 & 1 & 0.006 ** \\ 
   & \textbf{\textit{ChatGPT experience}} & 2.94 & 1 & 0.141 \\ 
\hline
\multirow{5}{*}{Fact-check}
   & \textbf{Visualization} & 20.16 & 3 & < .001 *** \\ 
   & \textbf{Collaboration} & 91.30 & 1 & < .001 *** \\ 
   & \textbf{Visualization × Collaboration} & 5.79 & 3 & 0.204 \\ 
   & \textbf{\textit{AI literacy}} & 0.20 & 1 & 0.651 \\ 
   & \textbf{\textit{ChatGPT experience}} & 0.21 & 1 & 0.651 \\ 
\hline
\multirow{5}{*}{Entire article}
   & \textbf{Visualization} & 7.11 & 3 & 0.114 \\ 
   & \textbf{Collaboration} & 262.05 & 1 & < .001 *** \\ 
   & \textbf{Visualization × Collaboration} & 2.81 & 3 & 0.423 \\ 
   & \textbf{\textit{AI literacy}} & 3.43 & 1 & 0.114 \\ 
   & \textbf{\textit{ChatGPT experience}} & 1.58 & 1 & 0.261 \\ 
\hline
\multirow{5}{*}{First draft}
   & \textbf{Visualization} & 0.76 & 3 & 0.860 \\ 
   & \textbf{Collaboration} & 247.03 & 1 & < .001 *** \\ 
   & \textbf{Visualization × Collaboration} & 8.56 & 3 & 0.089  \\ 
   & \textbf{\textit{AI literacy}} & 0.04 & 1 & 0.860 \\ 
   & \textbf{\textit{ChatGPT experience}} & 0.09 & 1 & 0.860 \\ 
\hline
\multirow{5}{*}{Style/clarity}
   & \textbf{Visualization} & 3.99 & 3 & 0.328 \\ 
   & \textbf{Collaboration} & 188.34 & 1 & < .001 *** \\ 
   & \textbf{Visualization × Collaboration} & 4.77 & 3 & 0.316 \\ 
   & \textbf{\textit{AI literacy}} & 2.39 & 1 & 0.305 \\ 
   & \textbf{\textit{ChatGPT experience}} & 0.28 & 1 & 0.598 \\ 
\hline
\multirow{5}{*}{Topic}
   & \textbf{Visualization} & 4.04 & 3 & 0.278 \\ 
   & \textbf{Collaboration} & 1.18 & 1 & 0.278 \\ 
   & \textbf{Visualization × Collaboration} & 6.53 & 3 & 0.148 \\ 
   & \textbf{\textit{AI literacy}} & 6.88 & 1 & 0.022 * \\ 
   & \textbf{\textit{ChatGPT experience}} & 8.16 & 1 & 0.021 * \\ 
\hline
\end{tabular}

\caption{Cumulative link mixed model results for perceived AI role with per-outcome FDR correction. $p$-values corrected within each outcome. Covariates are in italics.}
\label{tab:clmm_roles}
\end{table}

\begin{figure}
    \centering

    \begin{minipage}[b]{0.32\textwidth}
        \centering
        \includegraphics[width=\textwidth]{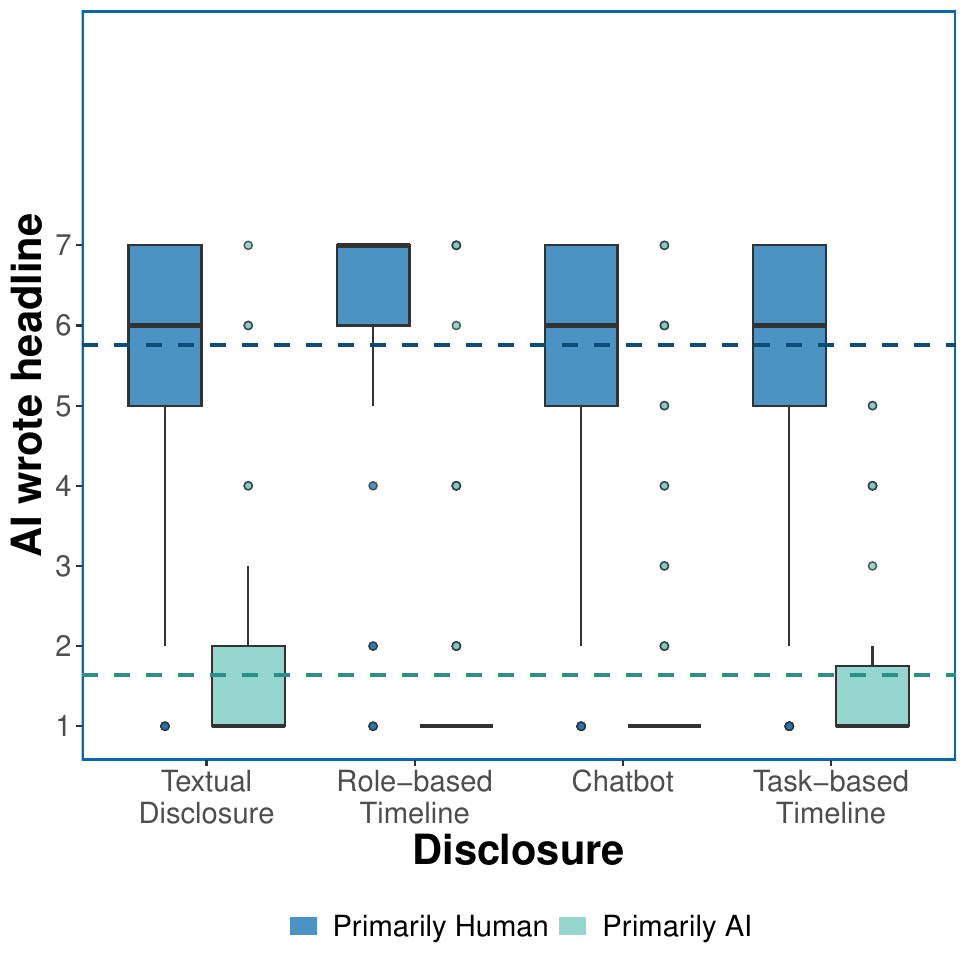}
        \caption*{(a) AI created the headline}
    \end{minipage}
    \hfill
    \begin{minipage}[b]{0.32\textwidth}
        \centering
        \includegraphics[width=\textwidth]{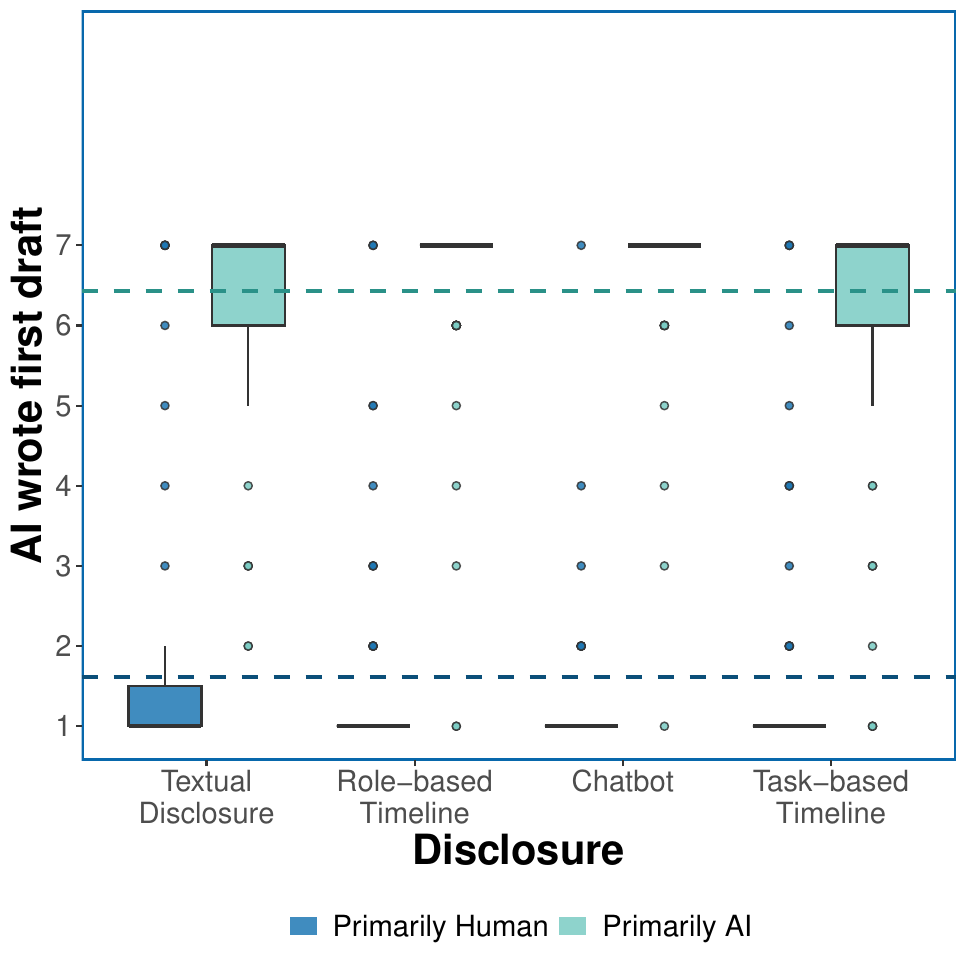}
        \caption*{(b) AI wrote a first draft of the article}
    \end{minipage}
    \hfill
    \begin{minipage}[b]{0.32\textwidth}
        \centering
        \includegraphics[width=\textwidth]{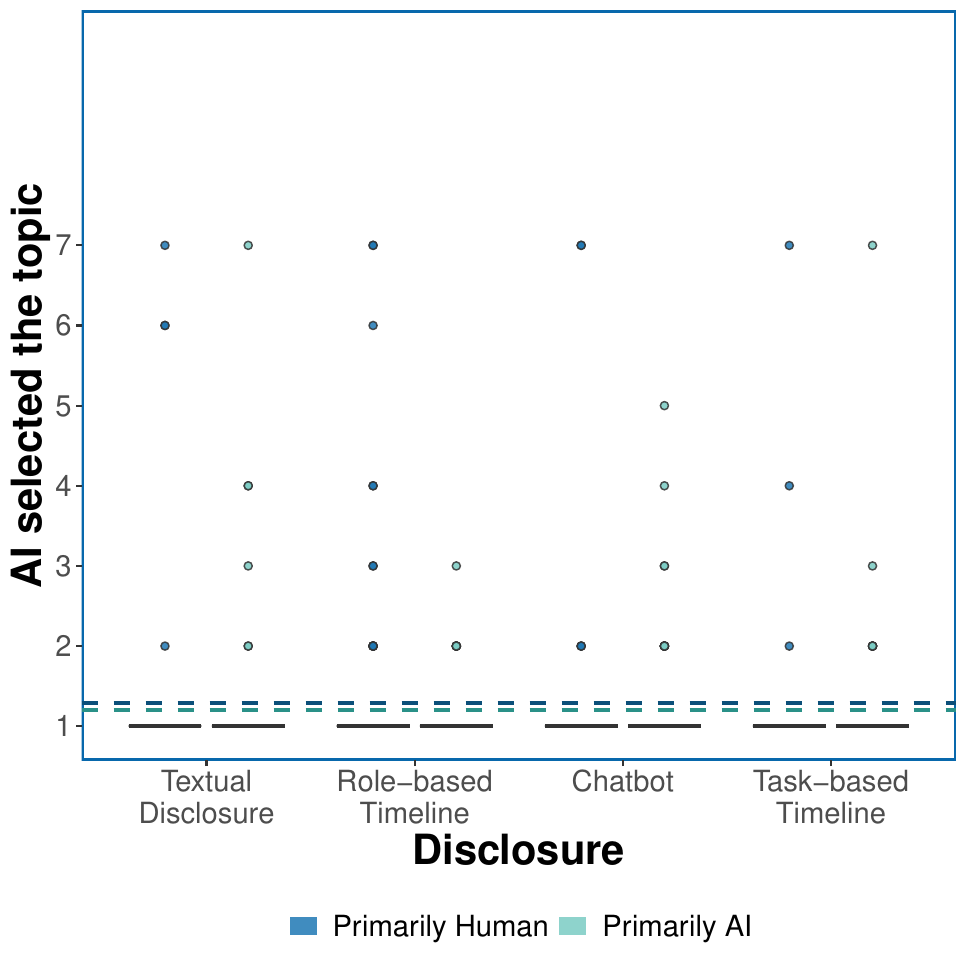}
        \caption*{(c) AI selected the topic of the article}
    \end{minipage}

    \vspace{0.1cm}

    \begin{minipage}[b]{0.32\textwidth}
        \centering
        \includegraphics[width=\textwidth]{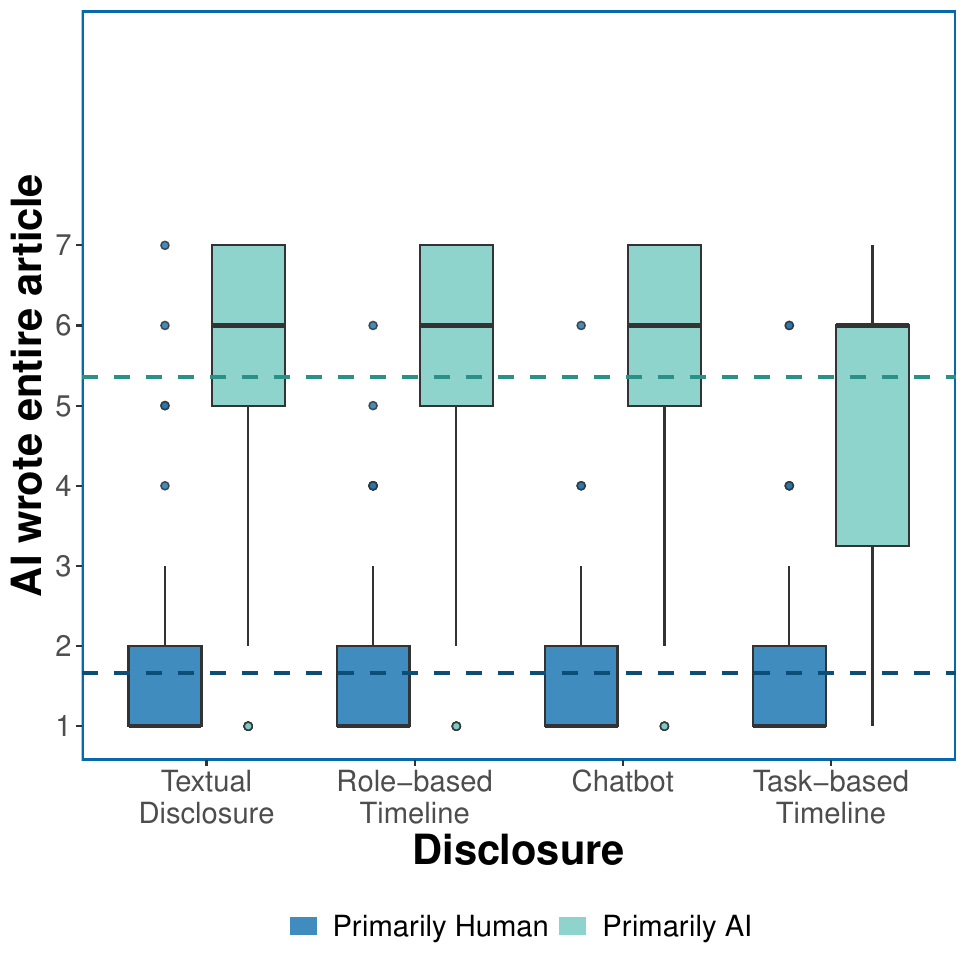}
        \caption*{(d) AI wrote the entire article}
    \end{minipage}
    \hfill
    \begin{minipage}[b]{0.32\textwidth}
        \centering
        \includegraphics[width=\textwidth]{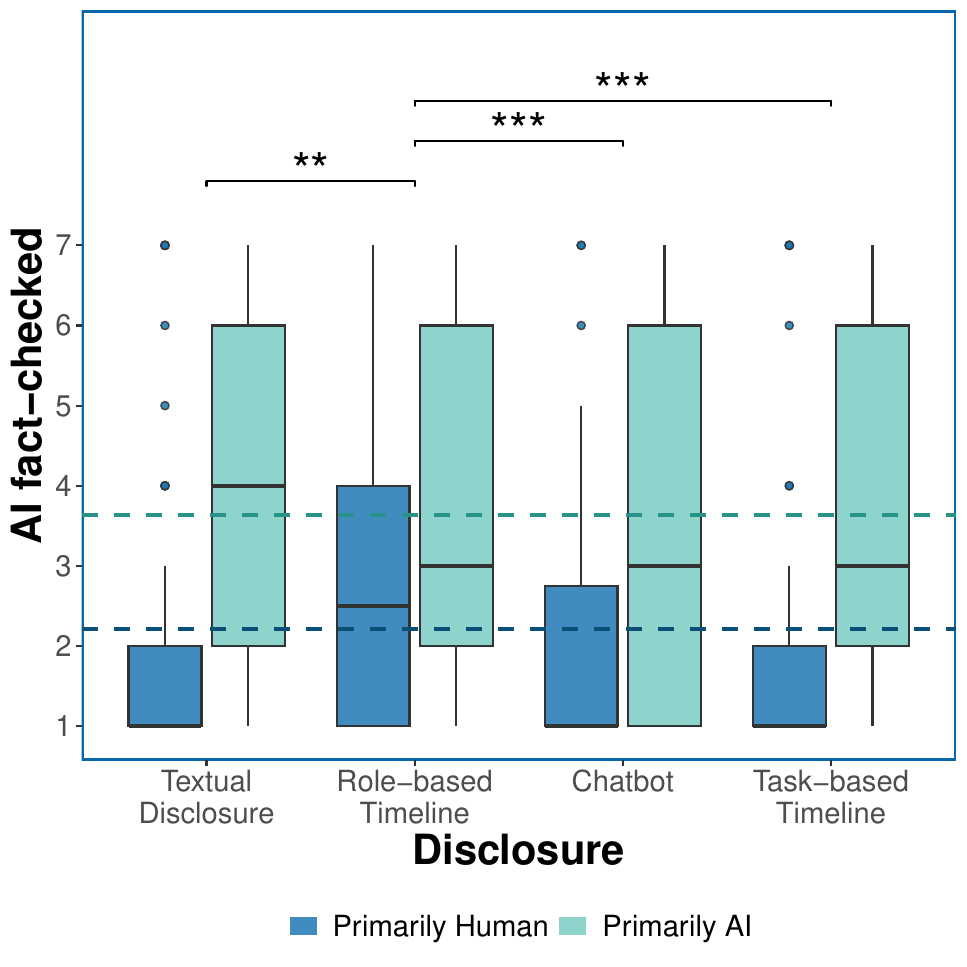}
        \caption*{(e) AI as used to fact-check the content}
    \end{minipage}
    \hfill
    \begin{minipage}[b]{0.32\textwidth}
        \centering
        \includegraphics[width=\textwidth]{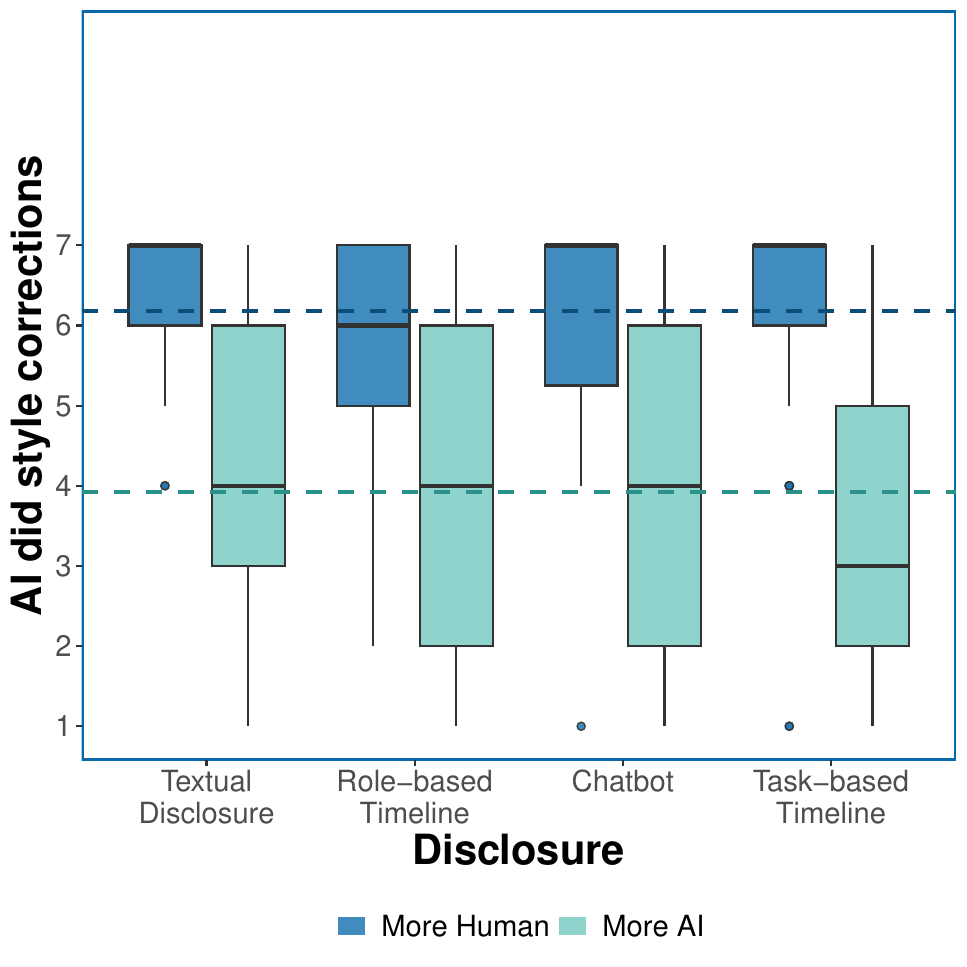}
        \caption*{(f) AI made stylistic changes}
    \end{minipage}
    
    \caption{Perceived AI involvement per role by visualization and collaboration ratio. Significance bar in black refer to significant main effect of visualization. Dashed line show average value per article version.}
    \label{fig:perceivedRolesAI}
    \Description{Five boxplots showing perceived AI involvement across roles, headline writing, drafting, topic selection, writing the full article, and fact-checking, by disclosure type and collaboration ratio. The X-axis lists disclosure visualizations, the Y-axis indicates perceived AI involvement on a scale from 1 to 7. Blue boxes represent Primarily Human articles, green boxes represent Primarily AI articles. Dashed horizontal lines mark the average rating per article version. Significance bars above plots indicate main effects or differences within AI or human conditions.}
    
\end{figure}

CLMMs revealed that the perceived role of AI can differed by disclosure visualization and that AI literacy and ChatGPT can also shape this (Fig.~\ref{fig:perceivedRolesAI}). Significant main effects on collaboration were expected due to the difference in responses for Primarily Human or AI. As seen in Table~\ref{tab:clmm_roles}, fact-checking showed a main effect of visualization ($p<.001$). Post-hoc tests revealed a main effect on visualization type. Across both contexts (Primarily Human and Primarily AI), Role-based Timeline made the perception of the role of AI in fact-checking more than Textual Disclosure ($\beta=-0.84, SE=0.25, z=-3.30, p=.002, OR(95\%CI)=0.43 (0.22, 0.84)$), Chatbot ($\beta=1.01, SE=0.26, z=3.95, p<.001, OR(95\%CI)=2.75 (1.40, 5.40)$) and Task-based Timeline ($\beta=0.91, SE=0.25, z=3.62, p<.001, OR(95\%CI)=2.48 (1.28, 4.80)$). For headline creation, higher AI literacy predicted lower perceived AI involvement ($OR(95\%CI)=0.78(0.66, 0.91), p=.006$). Similarly, for topic selection, higher AI literacy decreased perceived AI involvement ($OR(95\%CI)=0.53(0.33, 0.85), p=.022$), but more ChatGPT experience ($OR(95\%CI)=1.76(1.19, 2.60), p=.021$) increased the perceived contribution of AI in topic selection.
These findings suggest that disclosure visualizations can shape how readers attribute AI contributions to specific tasks (particularly fact-checking), and that the reader's AI literacy and ChatGPT experience plays a role in shaping these perceptions.

\subsubsection{Perceived clarity}
\begin{figure}
    \centering

    \begin{minipage}[b]{0.3\textwidth}
        \centering
        \includegraphics[width=\textwidth]{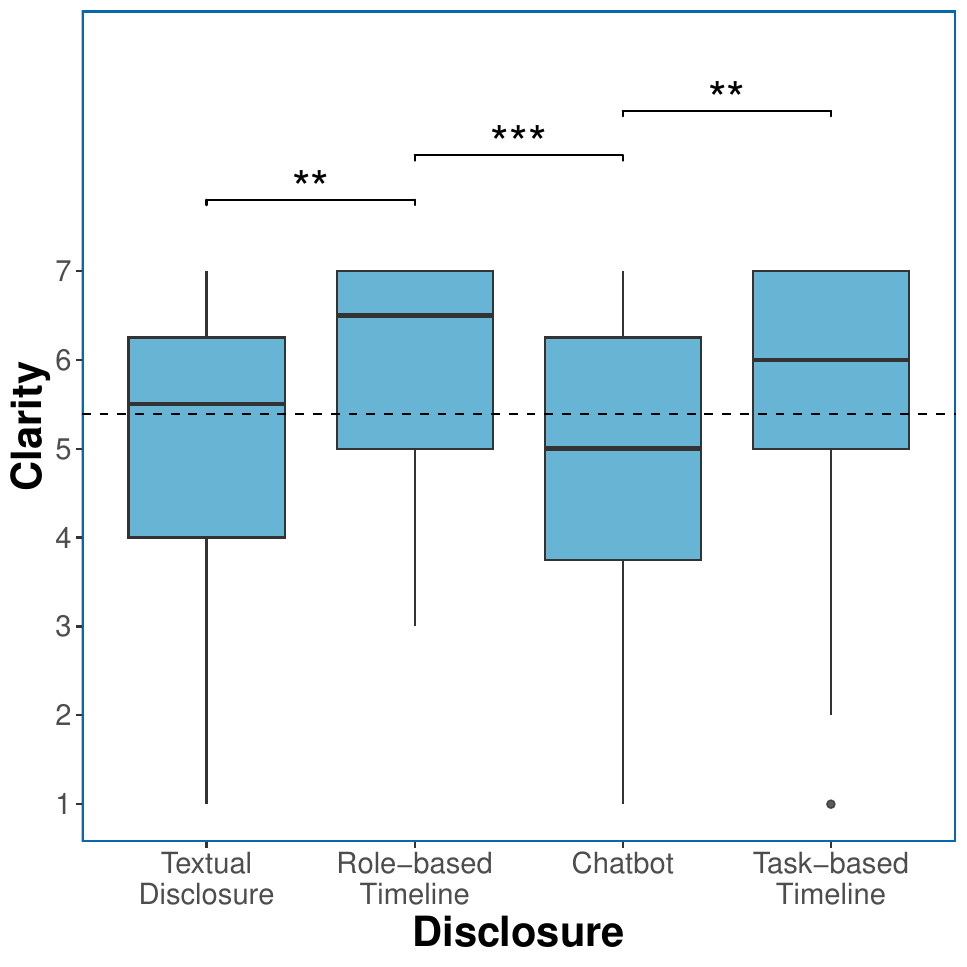}
        \caption*{(a) Perceived clarity}
    \end{minipage}
    \hfill
    \begin{minipage}[b]{0.3\textwidth}
        \centering
        \includegraphics[width=\textwidth]{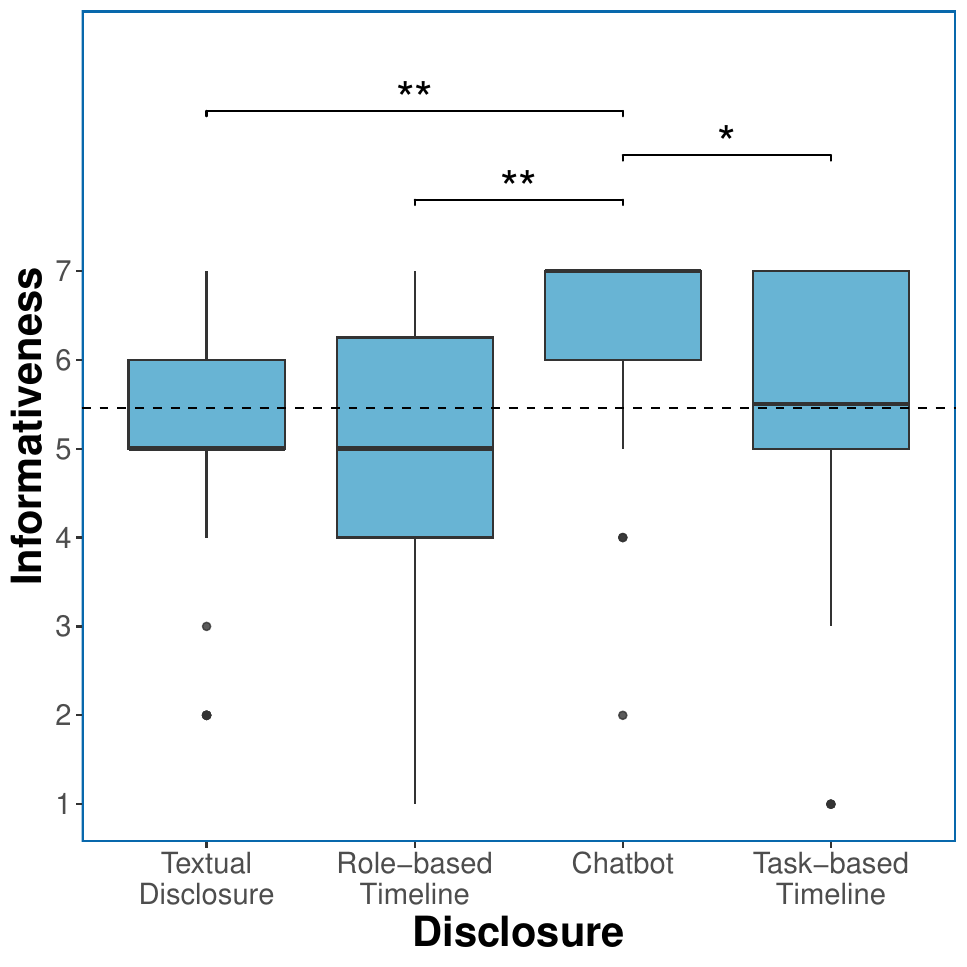}
        \caption*{(b) Perceived informativeness}
    \end{minipage}
    \hfill
    \begin{minipage}[b]{0.3\textwidth}
        \centering
        \includegraphics[width=\textwidth]{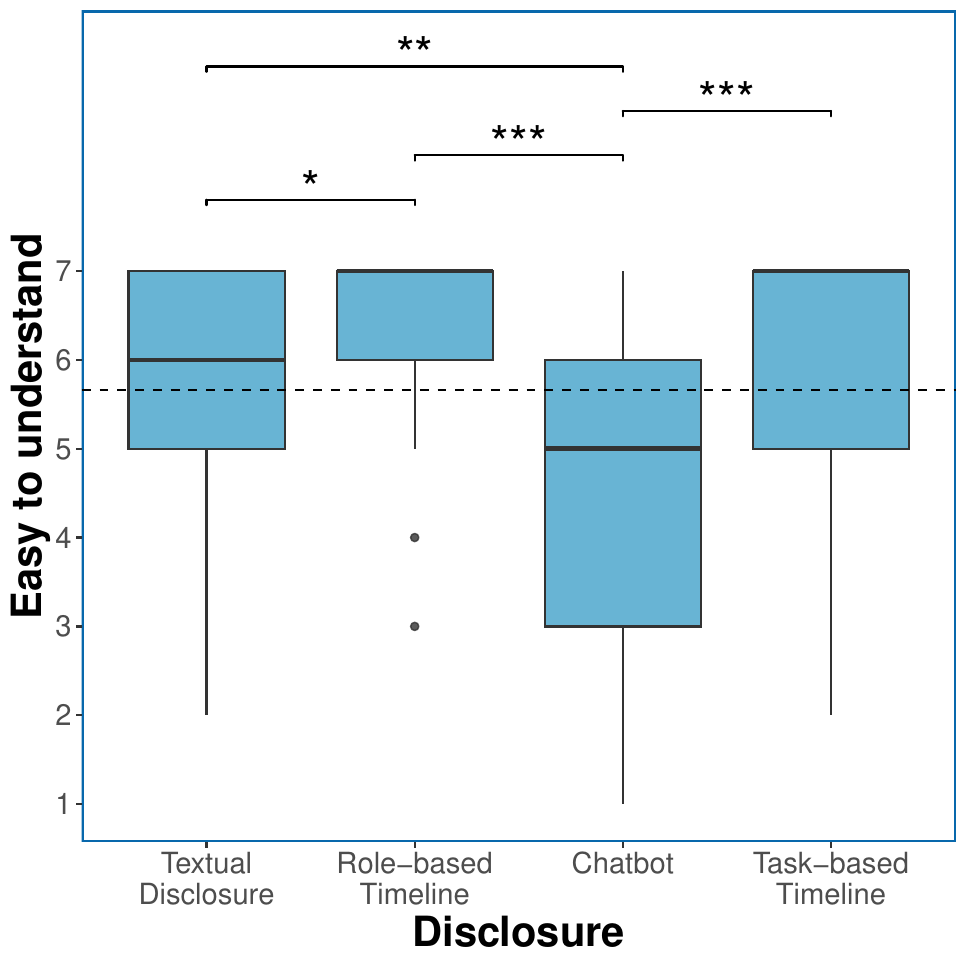}
        \caption*{(c) Perceived easiness to understand}
    \end{minipage}

    \vspace{0.3cm}

    \begin{minipage}[b]{0.3\textwidth}
        \centering
        \includegraphics[width=\textwidth]{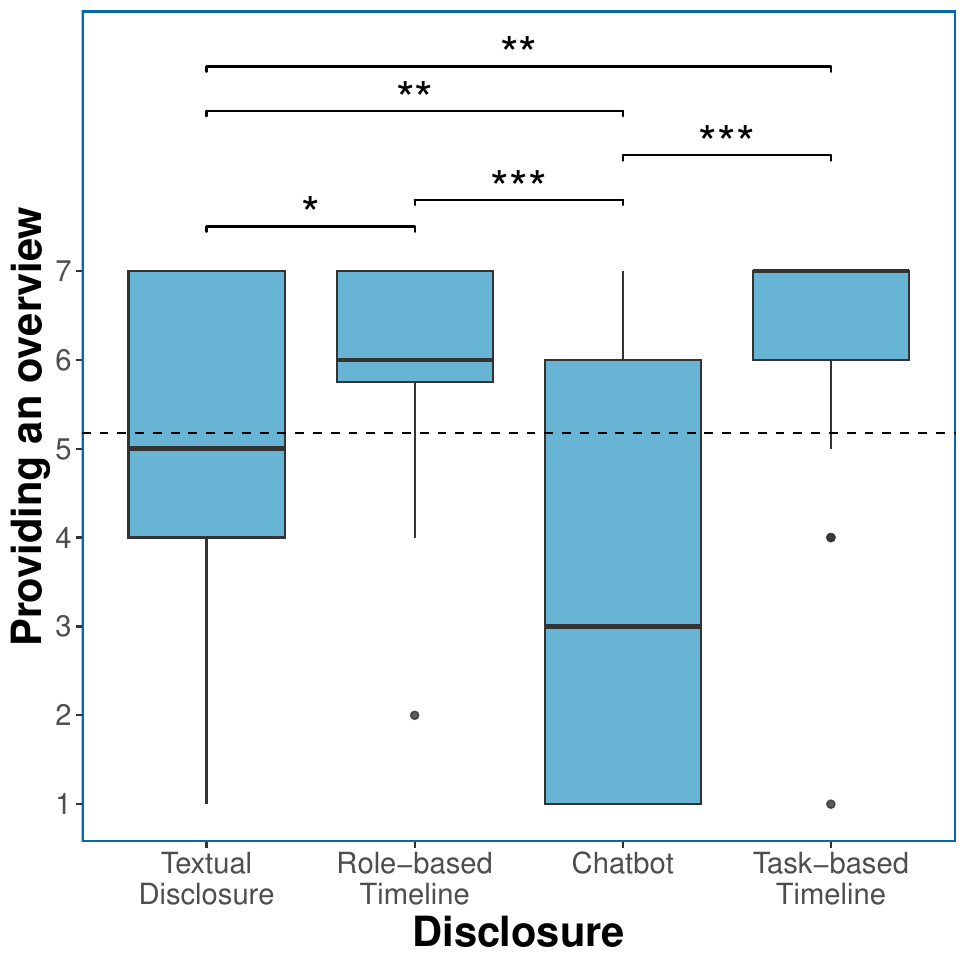}
        \caption*{(d) Providing an overview}
    \end{minipage}
    \hfill
    \begin{minipage}[b]{0.3\textwidth}
        \centering
        \includegraphics[width=\textwidth]{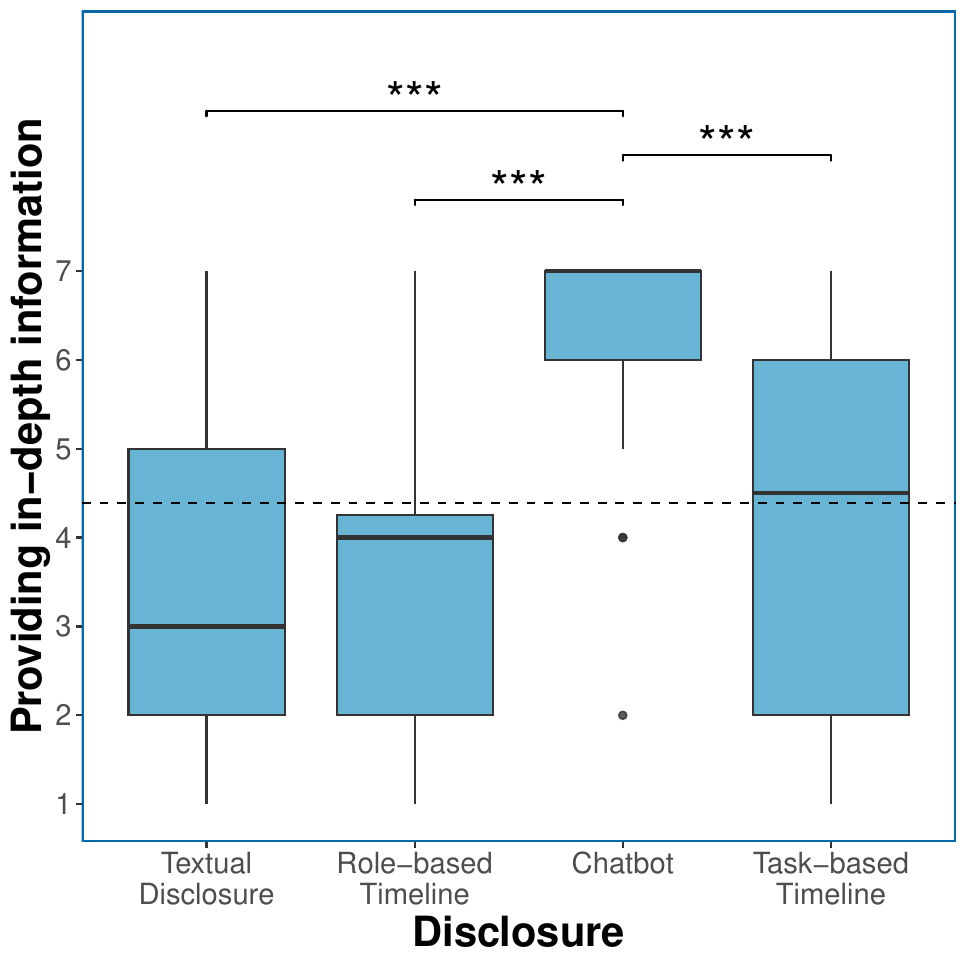}
        \caption*{(e) Providing in-depth information}
    \end{minipage}
    \hfill
    \begin{minipage}[b]{0.3\textwidth}
        \centering
        \includegraphics[width=\textwidth]{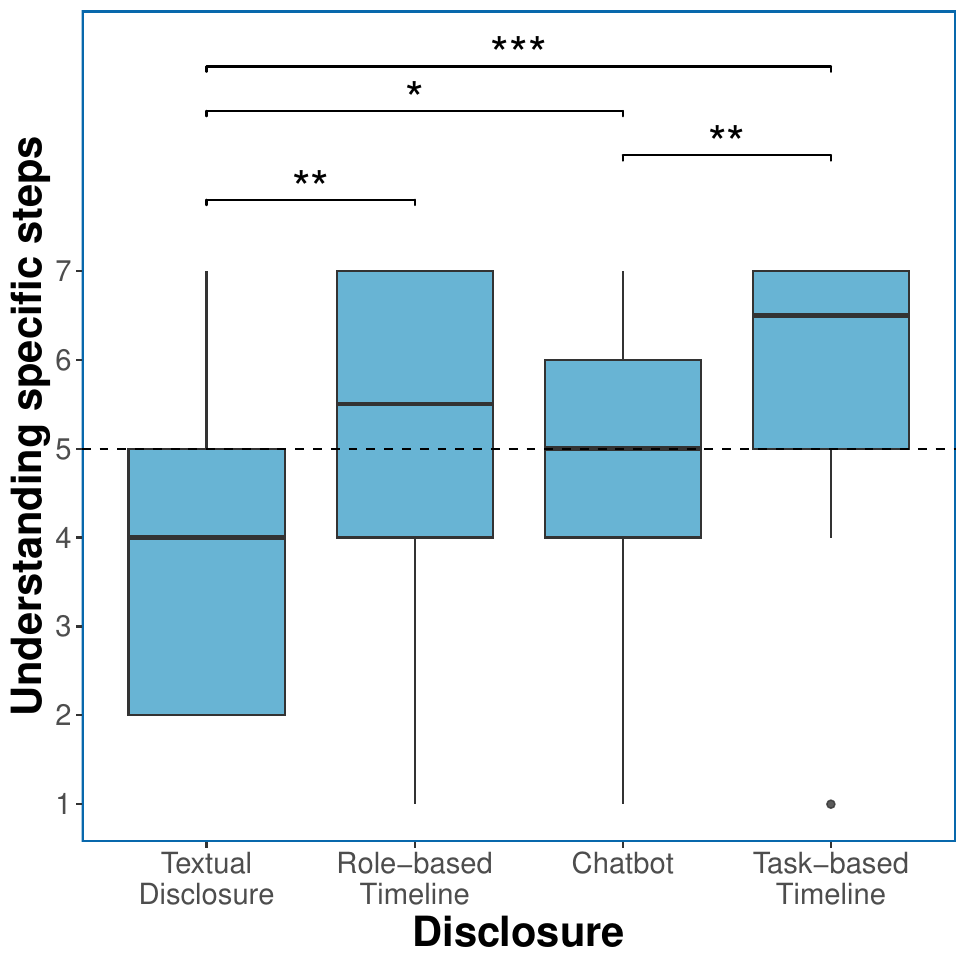}
        \caption*{(f) Understanding specific steps}
    \end{minipage}

    \vspace{0.3cm}

    \begin{minipage}[b]{0.3\textwidth}
        \centering
        \includegraphics[width=\textwidth]{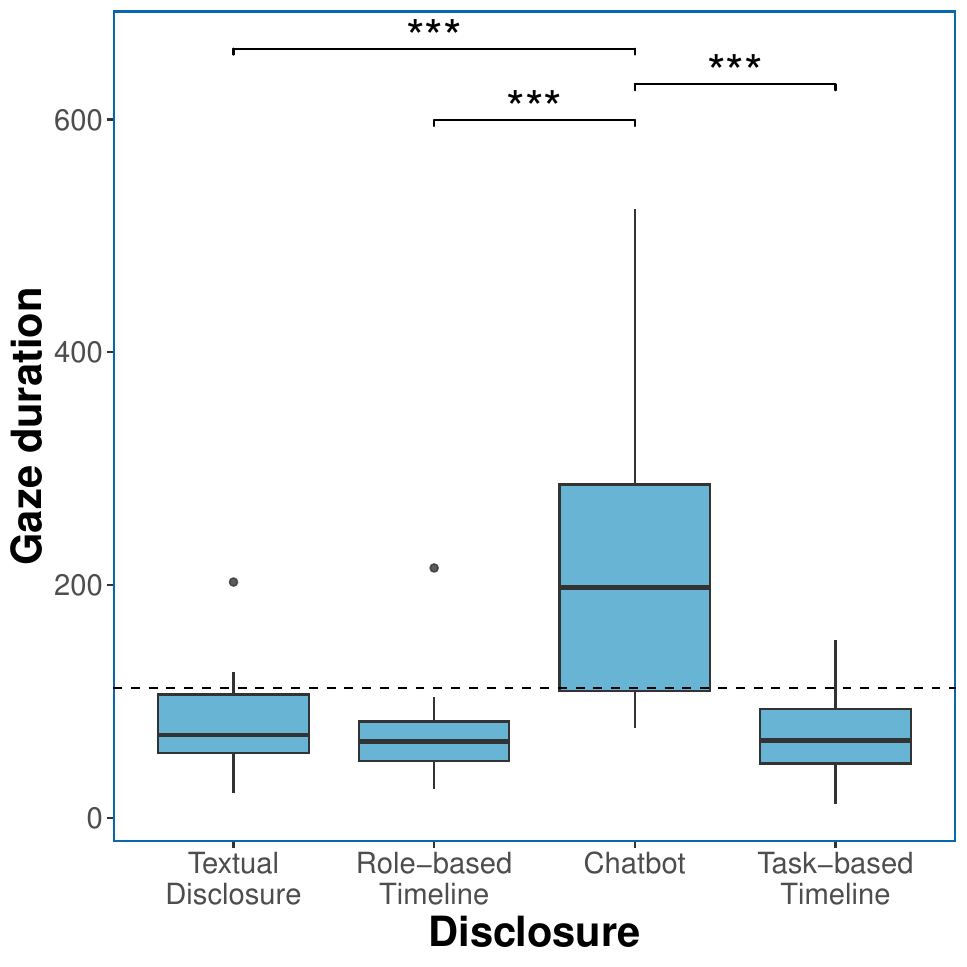}
        \caption*{(g) Gaze duration}
    \end{minipage}
    \hfill
    \begin{minipage}[b]{0.3\textwidth}
        \centering
        \includegraphics[width=\textwidth]{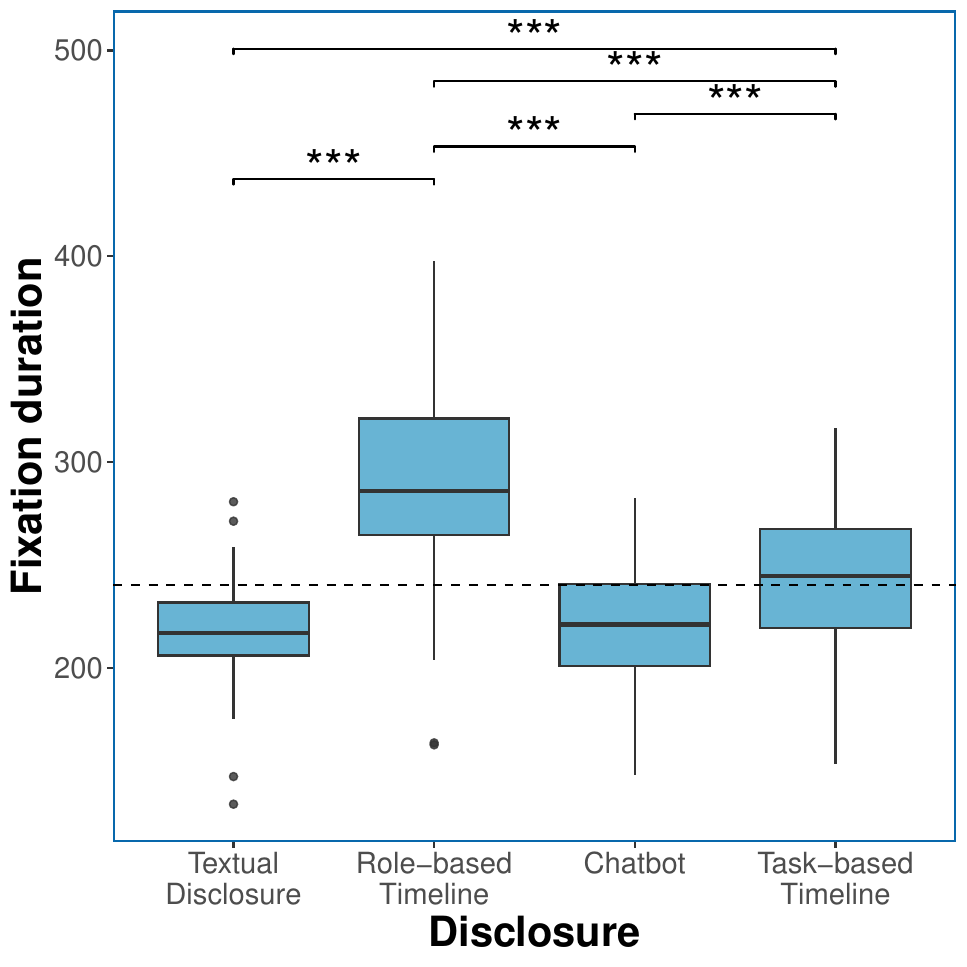}
        \caption*{(h) Fixation duration}
    \end{minipage}
    \hfill
    \begin{minipage}[b]{0.3\textwidth}
        \centering
        \includegraphics[width=\textwidth]{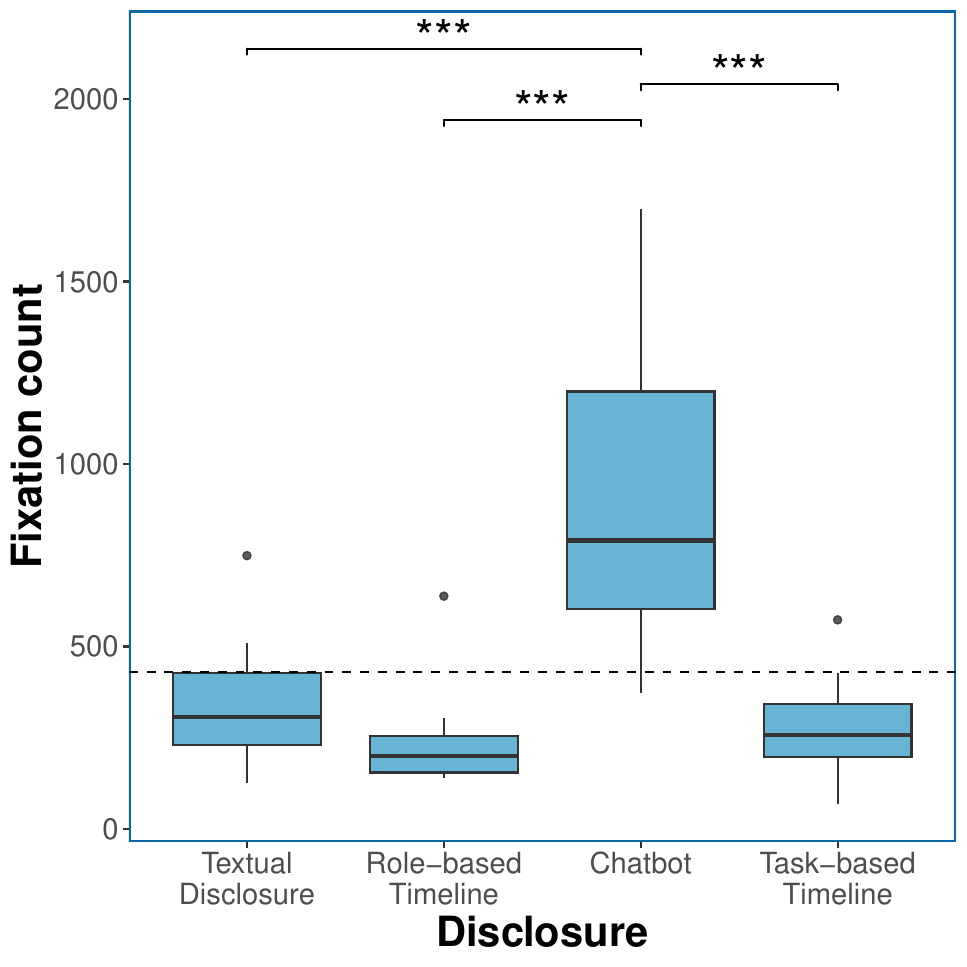}
        \caption*{(i) Fixation count}
    \end{minipage}

    \vspace{0.3cm}

    \begin{minipage}[b]{0.3\textwidth}
        \centering
        \includegraphics[width=\textwidth]{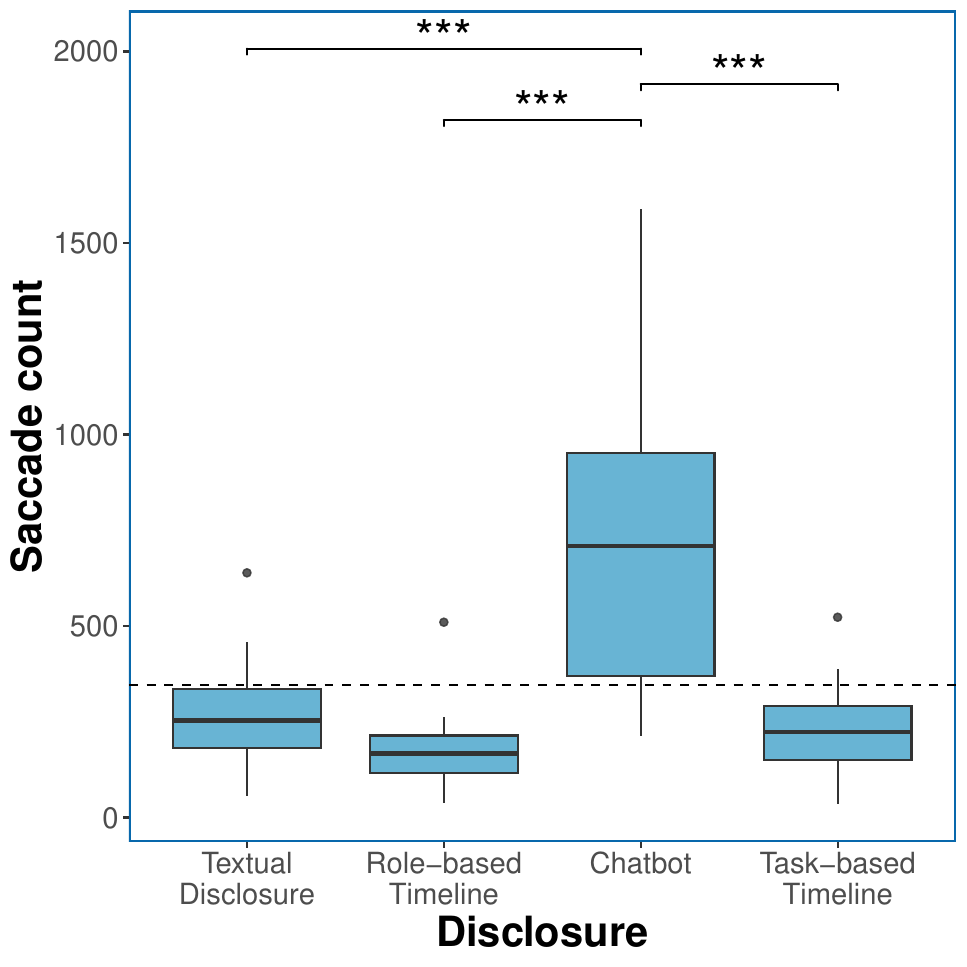}
        \caption*{(j) Saccade count}
    \end{minipage}
    \hfill
    \begin{minipage}[b]{0.3\textwidth}
        \centering
        \includegraphics[width=\textwidth]{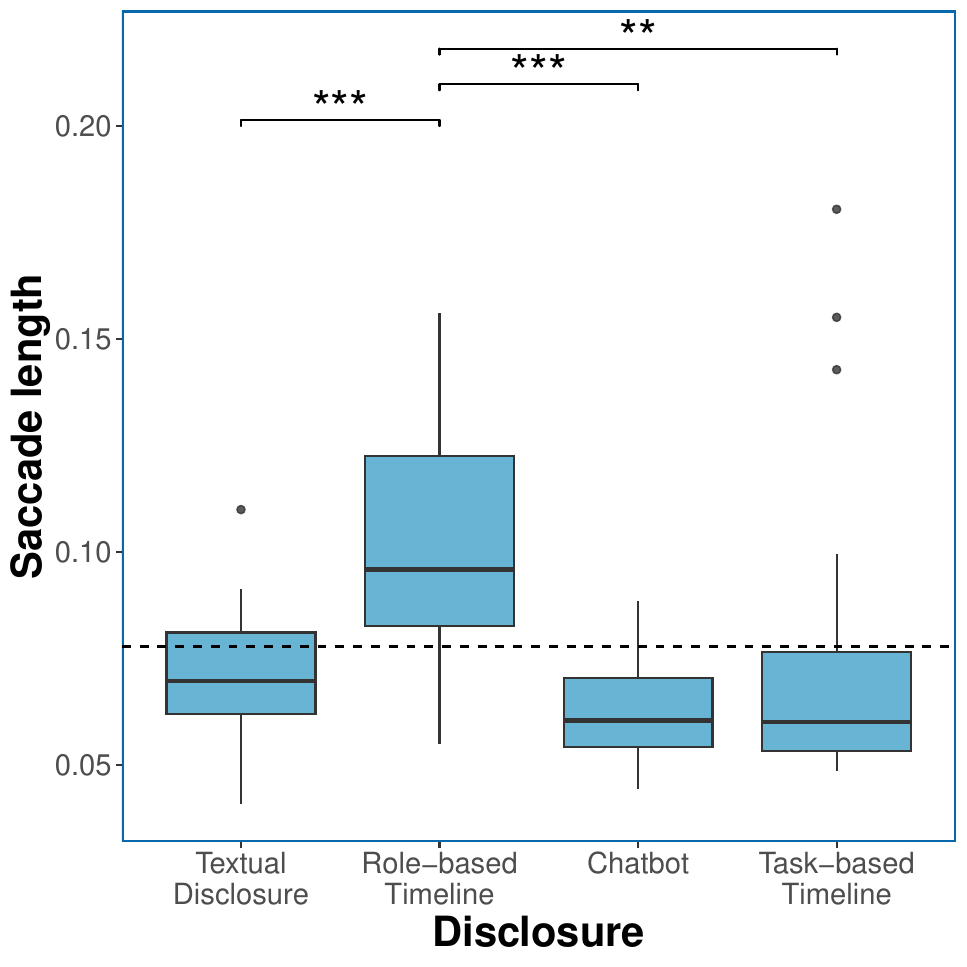}
        \caption*{(k) Saccade length}
    \end{minipage}

    \caption{Combined plots of response variables and gaze features per disclosure visualization.}
    \label{fig:combinedplots}
    \Description{Eleven boxplots showing participant ratings and eye-tracking measures across disclosure visualizations. The first six plots display subjective measures: perceived clarity, informativeness, ease of understanding, ability to provide an overview, in-depth information, and understanding specific steps. The last five plots show gaze features: gaze duration, fixation duration, fixation count, saccade count, and saccade length. Y-axes vary by measure, and dashed horizontal lines mark the mean. Significance bars above plots indicate differences between disclosure visualizations.}
\end{figure}

All visualizations were perceived clearly, with means above the midpoint. Fig.~\ref{fig:combinedplots} shows the distributions of all response variables and gaze features, with significance bars from CLMMs for response variables and LMMs for gaze features. There was a significant main effect of disclosure visualization on perceived clarity, , $p=.003$. Post hoc comparisons showed that both timelines were perceived most clear. Role-based Timeline was rated significantly clearer than Textual Disclosure ($\beta$=-1.55, $SE$=0.50, $z$=-3.08, $p$=.006, $OR(95\%CI)$=0.21 (0.06, 0.80)) and Chatbot ($\beta$=1.96, $SE$=0.52, $z$=3.79, $p$<.001, $OR(95\%CI)$=7.11 (1.81, 27.89)). Task-based Timeline was also rated clearer than Chatbot ($\beta$=-1.43, $SE$=0.50, $z$=-2.86, $p$=.008, $OR(95\%CI)$=0.24 (0.06, 0.89)).

\subsubsection{Perceived informativeness}
Disclosure visualization significantly affected perceived informativeness, $p=.018$. Post hoc comparisons showed that the Chatbot was rated as most informative compared to all other visualizations (TD: $\beta$=-1.64, $SE$=0.49, $z$=-3.35, $p$=.004, $OR(95\%CI)$=0.19 (0.05, 0.71), RT: $\beta$=-1.57, $SE$=0.49, $z$=-3.20, $p$=.004, $OR(95\%CI)$=0.21 (0.06, 0.76), TT: $\beta$=1.30, $SE$=0.49, $z$=2.63, $p$=.017, $OR(95\%CI)$=3.66 (0.99, 13.49)).

\subsubsection{Perceived easiness to understand}
A significant main effect of disclosure visualization was found for perceived easiness to understand, $p<.001$. Post hoc tests showed that the Chatbot was significantly less easy to understand than the other visualizations (Textual Disclosure: $\beta$=1.54, $SE$=0.49, $z$=3.13, $p$=.004, $OR(95\%CI)$=4.67 (1.27, 17.18), Role-based Timeline: $\beta$=2.89, $SE$=0.58, $z$=4.99, $p$<.001, $OR(95\%CI)$=17.97 (3.90, 82.79), and Task-based Timeline: $\beta$=-2.27, $SE$=0.54, $z$=-4.19, $p$<.001, $OR(95\%CI)$=0.10 (0.02, 0.43)). Additionally, the Role-based Timeline was significantly easier to understand than Textual Disclosure ($\beta$=-1.35, $SE$=0.52, $z$=-2.58, $p$=.015, $OR(95\%CI)$=0.26 (0.07, 1.03)). Differences between Role-based Timeline and Task-based Timeline were not significant.

\subsubsection{Providing an overview}
CLMM revealed a significant main effect of disclosure visualization, $p$ < 0.001. Post hoc comparisons showed that the other three visualizations were significantly better at providing an overview than the Chatbot: Role-based Timeline ($\beta$=2.82, $SE$=0.56, $z$=5.01, $p$<.001, $OR(95\%CI)$=16.70 (3.79, 73.53)), Task-based Timeline ($\beta$=-3.14, $SE$=0.58, $z$=-5.42, $p$<.001, $OR(95\%CI)$=0.04 (0.01, 0.20)), and Textual Disclosure ($\beta$=1.59, $SE$=0.51, $z$=3.14, $p$=.003, $OR(95\%CI)$=4.90 (1.29, 18.59)). Furthermore, the Role-based and Task-based Timeline provided a significantly better overview than Textual Disclosure (RT: $\beta$=-1.23, $SE$=0.49, $z$=-2.50, $p$=.015, $OR(95\%CI)$=0.29 (0.08, 1.07), TT: $\beta$=-1.55, $SE$=0.51, $z$=-3.05, $p$=.003, $OR(95\%CI)$=0.21 (0.06, 0.81)).

\subsubsection{Providing in-depth information}\label{sec:indepth}
There is a significant main effect of disclosure visualization $p$ < 0.001. Post hoc comparisons showed that the Chatbot visualization was significantly better at providing in-depth information compared to the other three visualizations: Textual Disclosure ($\beta$=-3.58, $SE$=0.60, $z$=-5.99, $p$<.001, $OR(95\%CI)$=0.03 (0.01, 0.13)), Role-based Timeline ($\beta$=-3.49, $SE$=0.59, $z$=-5.88, $p$<.001, $OR(95\%CI)$=0.03 (0.01, 0.15)), and Task-based Timeline ($\beta$=2.95, $SE$=0.58, $z$=5.10, $p$<.001, $OR(95\%CI)$=19.12 (4.15, 88.03)).

\subsubsection{Providing an understanding of specific steps in the process}
CLMM results for providing an understanding in specific steps in the process also revealed a significant main effect of visualization  type $p$ < .001. Post hoc comparisons showed that the Textual Disclosure performed significantly worse than all other visualizations at showing the specific steps: Role-based Timeline ($\beta$=-1.65, $SE$=0.48, $z$=-3.42, $p$=.002, $OR(95\%CI)$=0.19 (0.05, 0.69)), Chatbot ($\beta$=-1.05, $SE$=0.45, $z$=-2.33, $p$=.030, $OR(95\%CI)$=0.35 (0.11, 1.15)), and Task-based Timeline ($\beta$=-2.54, $SE$=0.51, $z$=-5.01, $p$<.001, $OR(95\%CI)$=0.08 (0.02, 0.30)). Additionally, the Task-based Timeline was significantly better at showing the specific steps than the Chatbot ($\beta$=-1.50, $SE$=0.48, $z$=-3.09, $p$=.004, $OR(95\%CI)$=0.22 (0.06, 0.80)).

\subsubsection{Gaze duration}
A linear mixed-effects model showed a significant main effect of disclosure visualization, F(3,57) = 31.4, $p$<0.001, partial $\eta^2$=.62. Post hoc comparisons showed that the Chatbot had significantly longer gaze durations than all other visualizations (all $p$<0.0001), with very large effect sizes (vs. RT: $d$=2.58, vs. TD: $d$=2.41, vs. TT: $d$=2.52). Which is in line with the time spent on the surveys (Sec.~\ref{sect:meantime}), and with the expectations as the Chatbot required more interactions. Additionally, individuals with higher AI literacy score had longer gaze durations ($\beta=23.57, p=.019$), while users with more ChatGPT experience had shorter gaze durations ($\beta=-18.95, p=.013$), as more experience requires less time to process information.

\subsubsection{Fixation duration}
A significant main effect of disclosure visualization was also found for fixation duration, F(3,57) = 45.96, $p$ < 0.001, partial $\eta^2$=.71. Participants had significantly longer fixations for Role-based Timeline than all other types (all $p$ < 0.001, vs. TD: $d$=3.27, vs. C: $d$=3.15, vs. TT: $d$=1.99). The fixations on Task-based Timeline were also significantly longer compared to Textual Disclosure ($p$ < 0.001, $d$ = 1.28), and Chatbot  ($p$ < 0.001, $d$ = 1.16). The Chatbot did not differ significantly from Textual Disclosure. Participants with more experience in using ChatGPT had a significantly shorter fixation duration ($\beta=-19.45, p=.002$).

\subsubsection{Fixation count}
There was a significant main effect of disclosure visualization for fixation count, F(3, 57) = 44.85, $p$ < 0.001, partial $\eta^2$=.70. The Chatbot had significantly more fixations compared to all other visualizations (all $p$ < 0.001, vs. TD: $d$=2.69, vs. RT: $d$=3.22, vs. TT: $d$=2.98). The fixation count was also influenced by the AI literacy of the participants, as a higher literacy had an increase in fixation count ($\beta=69.26, p=.021$).

\subsubsection{Saccade count}
For saccade count there was also a significant main effect found, F(3, 57) = 34.44, $p$ < 0.001, partial $\eta^2$=.64. Similar to fixation count, saccade counts were significantly higher for Chatbot than all others (all $p$ < 0.001, vs. TD: $d$=2.33, vs. RT: $d$=2.85, vs. TT: $d$=2.59). For this feature, both ChatGPT experience and AI literacy were significant covariates. AI literacy increased the saccade count ($\beta=66.05, p=.046$), while ChatGPT experience decreased this count ($\beta=-49.89, p=.045$).

\subsubsection{Saccade length}
There was a significant main effect of disclosure visualization, F(3, 57) = 11.11, $p$ < 0.001, partial $\eta^2$=.37. Post hoc test showed that Role-based Timeline had significantly longer saccades than all other visualizations (vs. TD: $p$ < 0.001, $d$ = 1.33,  vs. TT: $p$ = 0.003, $d$ = 1.06, vs. C: $p$ < 0.001, $d$ = 1.75).

\subsection{Qualitative results}
We analyzed the data with deductive thematic analysis \cite{braun2012thematic}. Two authors separately analyzed the qualitative data before meeting to discuss our analysis. First, we coded it according to the interview questions. Then within each topic, we analyzed emerging themes. Recurring ideas across participants were noted down, but due to the semi-structured nature of the interview, it is important to mention that if a participant did not mention something, it does not mean they disagree. Participants are labeled P1-P32. Four themes emerged: initial interpretations of the visualizations, preferences for different designs, reflections on interactivity and information structure, and disclosures in real-world news contexts.

\subsubsection{Interpretation of the visualizations}\label{qual:interpretation}
The initial reaction of participants on the disclosures was overall positive. They were described as “\textit{useful and helpful for the average consumer reading the news}” [P19], and “\textit{illustrative, functional}” [P30]. Three participants mentioned they had never seen these kinds of visualizations before “\textit{I hadn't really seen those before, so I was kind of surprised by them. And I was just curious to see what was done by journalists and what was done by an AI tool}” [P14]. Four participants appreciated the transparency “\textit{I think it's nice to be transparent about AI use}” [P4].

Participants also reflected on how the visualizations shaped their understanding of how the article was created. Seven said they could not initially tell AI was used “\textit{I didn't really have an idea on what was done by AI and what was done by the journalist}” [P14]. For six others, the visualizations matched their expectations “\textit{this is written by AI and then the visualization confirmed it for me}” [P17]. Additionally, two participants mentioned the visualizations created more nuance “\textit{It still made me change from the opinion that this was an AI generated article to actually a blend of AI and human generated article}” [P16]. However, two other participants also expressed some skepticism “\textit{if it's stated like a journalist wrote an article, I'm not sure whether to believe that or not}” [P1], “\textit{I feel like some parts were not true. Maybe it's just the wording in it}” [P9].

\subsubsection{User preferences per disclosure visualization}\label{qual:preference}
Overall, preferences were divided, but clear patterns emerged. Task-based Timeline was most frequently ranked first, described as “\textit{the best combination of informative, in-depth and visually clear as well. So it immediately provides you an overview}” [P25]. Yet, nine participants overlooked the hover interaction, with several stating they “\textit{didn't know [it] was interactive}” [P5]. Some disliked the interaction entirely, “\textit{I didn't like having to hover. If you're going to do a visualization, then just show it directly}” [P7]. Suggestions included changing the element to “\textit{a small text, above or below the circles, that you could click if you wanted to dive deeper}” [P25]. Role-based Timeline was also frequently favored, as it was clear and easy to understand: “\textit{the level of AI against human input is clearly stated here}” [P32], and “\textit{the pictures make it more easy to understand}” [P27]. Participants also liked that “\textit{it gives a good overview and it's to the point, you don't have to hover}” [P20]. Still, others found it less detailed than the task-based version, “\textit{it did not provide as much detail as the task-based timeline}” [P16], and potentially misleading, “\textit{it gives you the idea that the roles are done half half, then you need to read it}” [P12].  

The Chatbot was described as most informative, valued for depth and transparency: “\textit{I liked the ability to interact with it ... it could explain what AI can do. It has so many possibilities to really go in depth, and especially with journalism, I'd like it to be very transparent and adherent to integrity}” [P11]. It was also appreciated for giving control: “\textit{it was nice to be given the option to consume the information that you are interested in}” [P31]. In contrast, it was often critiqued as overwhelming, “\textit{a bit too much information. I’m not sure if people are really going to read this}” [P8], and some struggled to navigate between questions. The Chatbot was seen as clear and novel, but lacking an overview.  The Textual Disclosure was most often ranked lowest, described as “\textit{too textual, looks like a legal disclaimer rather than an informative thing for users}” [P10]. Others noted it “\textit{didn’t grab my attention}” [P11] and was “\textit{not very detailed}” [P16]. Still, seven participants valued its simplicity, “\textit{it has a short overview of all the information needed}” [P14], appreciating that it was easy to understand.

\subsubsection{Interaction style and information structure}\label{qual:style-structure}
Preferences reflected broader design considerations: interaction style (static vs. interactive) and information structure (role-based vs. task-based). Nine participants favored interactive disclosures for exploration and engagement, e.g., “\textit{I understand the process better when it's interactive because you actively click on something because you want to know what happened}” [P3], and control of information was appreciated “\textit{you can look for more information if you want it}” [P15]. In contrast, six participants favored the static disclosures for the overview and completeness. Some doubted they would interact with it on a news page “\textit{I don't think it's really used if it would be in an article}” [P5] and liked having “\textit{all the information already printed out in a schematic way. So it's good that you don't need to even move a finger to get other information}” [P10].

Participants were also divided in their preferences for representing information as role or task-based. Fourteen participants preferred role-based structures, “\textit{I was more concerned about the roles since I already knew the steps}” [P10]. On the other hand, eleven participants preferred the task-based structures “\textit{I felt the process depicts a timeline better}” [P31]. Remarkably, familiarity with the journalistic workflow shaped preferences. A participant with writing experience preferred task-based information: “\textit{Because I write articles. So then I want to know what they did and every step}” [P15]. This suggests that disclosure visualizations should account for varying levels of domain familiarity among news readers.

\subsubsection{AI disclosures in real-world news}\label{qual:realworld}
Participants reflected on how these disclosures might influence their engagement with news. Five noted the impact would depend on the topic, especially political or societal issues: “\textit{then I think it's really important some things are properly fact checked. And then I would want to know who did what}” [P11]. Thirteen emphasized disclosures as more important in high-impact contexts, “\textit{especially who has done the research}” [P8] and “\textit{the specific steps in which AI is involved become way more important}” [P25]. Nine participants wanted more detail about article creation, for example: “\textit{if it's more high-impact... then I would want to know if AI made that assumption or a person}” [P30]. Two stated they would not read AI-generated news at all: “\textit{I wouldn't want to read an article that would be partly created by AI if it would be a critical article. Because I think you should think about the impact that an article can make. So then I think it is more wise to only have a journalist write it}” [P5]. In contrast, two participants did not care about AI usage: “\textit{I don't really want to know if the AI did it or the human did it. I don't care actually}” [P2]. Two others said they would scan more quickly if they knew an article was AI-written. At the same time, eight said disclosures would make them read more critically, “\textit{if you see how much AI has done... you might become a bit more critical or pay more attention}” [P21]. Two others noted the visualizations gave them a better sense of AI’s capabilities: “\textit{it made me realize that AI could do a lot more than I thought}” [P18].  

Fifteen participants would like to see such visualizations on news platforms, as “\textit{transparency is really important}” [P17]. However, five expressed skepticism, they mentioned “\textit{It's like a romantic idea of journalism and investigative journalism still being very much primarily human based. If I were to see an AI label, I’d probably become skeptical}” [P11]. Two said they might even stop reading outlets that disclosed AI usage: “\textit{I would steer more towards other platforms or news sites if I knew that that platform uses AI}” [P5].  Placement preferences were divided. Fifteen wanted disclosures upfront for ethical transparency, e.g., “\textit{I think it's more ethical to make people aware as soon as possible}” [P11]. Eleven preferred them at the bottom to avoid distraction: “\textit{I think it takes away from the article if you immediately show this was generated with AI}” [P30]. A few suggested optional or hybrid approaches, such as “\textit{a small icon above the article and then the full visualization under it}” [P25], or side placement “\textit{just because sometimes maybe you don't finish the whole article}” [P4].

bl\section{Discussion}

\subsection{How can we visually represent the ratios of human-AI collaboration output in news production and beyond?} 

A primary goal of this research (\textbf{RQ1}) was to explore how human–AI collaboration in news production can be effectively communicated to readers through visual disclosures. To address this, we drew on principles from HCI and Information Visualization, and involved designers and HCI experts in co-design sessions. These sessions generated 69 design ideas, spanning both static and interactive formats, and explored a wide design space of how to represent human-AI collaboration ratios. Despite this broad scope, we found two overarching strategies emerged: ratio-focused and process-focused. Ratio-focused designs emphasized proportions through familiar visual methods such as pie charts, bar charts, percentages, or stamps, echoing common ratio visualizations used in media \cite{siirtola2014bars,Bianconi2024}. Process-focused designs instead emphasized the steps taken by journalists or AI, aligning with \citet{wittenberg2024Labeling}'s framework for process-based disclosures. While these approaches reveal complementary ways of framing human–AI collaboration for readers, a key challenge across designs was how to communicate “who did what” without overwhelming readers with detail, a point reminiscent of how consent banners (e.g., cookies for accepting General Data Protection Act) can exhibit dark patterns \cite{Gray2021darkpatternsconsentbanners, nouwens2020dark}, with users getting habituated to such consent interfaces \cite{Habib2022consentcookieinterfaces}. As such, designers in our study explored varying levels of information depth, from simple overview visualizations to multi-layered designs, drawing on principles of progressive disclosure \cite{springer2020progressive}. Furthermore, this draws on \citet{burrus2024unmasking}'s approach, who argue that readers should be informed without overwhelming them with technical details. While we find that there is no one-size-fits-all solution to visualizing human–AI collaboration in news articles, we find that such visualizations must balance: (1) proportional representation of human and AI contributions, (2) process-based explanations of how those contributions were made, and (3) levels of information detail that can support readers with different engagement goals and AI literacy levels.

Taken together, we contribute an initial design space where we consider nuanced approaches in AI system disclosure design, with the objective of clearly communicating human–AI collaboration. This is to ensure readers to make more informed judgments about credibility and accountability. While our scope was limited to these designs, our work raises questions about optimal disclosure placement (cf., \cite{longdisclosure}), layering, and device as well platform-based disclosure adaptation, since for example disclosures that work on a desktop site may not easily translate to mobile or social media feeds. Furthermore, beyond textual content, these early visualizations provide a starting point toward adaptation to other modality-dependent approaches, for example provenance indicators for AI-edited images \cite{Trattner2025}, timeline visualizations for video production \cite{ide2025signalsprovenancepractices}, or even metadata encryption in audio files \cite{TransparentMeta2025}. We believe our work serves as a key starting point for ensuring more nuanced disclosure design in this era of blended work \cite{Constantinides2025futureofworkblended}, within journalism and beyond. This includes key domains such as science communication or within educational activities, where transparency about human–AI collaboration is equally important \cite{weingart2016sciencetrust, Guenther09052025sciencejournalist,zhang2025educationtrust}. For example, similarly to how journalists would carefully disclose their usage in a given format, we believe the same process can be applied towards AI usage disclosure for academic manuscripts, medical practitioners, and possibly any domain where AI usage is (a) composed of tasks with differing influences of AI contributions (b) (eventually) governed by the need for disclosure (e.g., through policy or best practices). In a future where combinations of humans and AI (cf., \cite{vaccaro2024combinations}) will increasing become a de facto standard, our work takes a leap toward envisioning a media ecosystem where audiences (whether news consumers or otherwise) are not only told that AI was involved, but can meaningfully understand and evaluate its role, in a usable manner. 



\subsection{How do different disclosure visualizations communicate human–AI collaboration to readers?}

We evaluated four human–AI collaboration disclosure visualizations on user perception (\textbf{RQ2}) in a lab-based study combining eye tracking, questionnaires, and interviews. We found that participants distinguished between high and low AI contributions across all designs (Fig.~\ref{fig:perceived_HAICollab_barplot}), validating disclosures as effective tools to communicate collaboration dynamics. This echoes Zier \& Diakopoulos \cite{zierlabeling}, who state that showing both human and AI involvement is necessary to demystify assumptions about AI. Overall, all visualizations were rated positively, but each revealed distinct strengths suited for certain purposes or contexts. Surprisingly, the visualization format itself influenced how readers attributed human and AI roles. For primarily AI-written articles, the Task-based Timeline made readers see greater human involvement for writing the entire article, while Role-based Timelines shifted primarily human-written articles towards the perception of more AI contribution. At specific role level, only fact-checking was perceived with more AI involvement due to the Role-based Timeline, in both context. These amplifications show that disclosure designs (likely in leading users to focus on different aspects of the information and visual design) do not just communicate information but can actively alter the balance of human and AI contribution understanding, depending on how roles and tasks are represented. 

Additionally, AI literacy and ChatGPT experience influenced perception of human-AI collaboration and visual engagement. Individuals with higher AI literacy demonstrated longer gaze durations, and more fixations and saccades, but perceived less AI involvement to headline creation and topic selection. This indicates to a more critical evaluation. However, more experienced ChatGPT users spent shorter time, and had fewer and shorter fixations and saccades, but attributed more AI involvement to topic selection. These findings reveal that the communication of human-AI collaboration by a visualization also depends on the readers' experience and literacy. Below we list findings related to each of our visualizations:

\textbf{Textual Disclosure} is a more collaborative focused resemblance of current platform labels, but was often described as too general or overlooked. Eye tracking confirmed this with shorter fixations suggesting scanning rather than deep processing. Participants described it as “too textual” or “like a legal disclaimer” (Section~\ref{qual:preference}), which helps explain why it was least effective at conveying workflow steps or providing an overview. Additionally, it highlights the importance of Epstein et al.’s \cite{epstein_fang_arechar_rand_2023} observation that the language used in labels has an immense influence on user perceptions. Although a common format nowadays, our findings suggest that it is not the most effective in communicating human-AI collaboration. 

\textbf{Role-based Timeline} provided a significantly better understanding of the steps, also without requiring any user interaction. Participants valued its static format for quick scanning, and eye tracking data indicated longer fixations and broader eye movements. Both timelines were perceived significantly clearer than the Chatbot and Textual Disclosure, and the Role-based Timeline was also seen as easier to understand than Textual Disclosure, which was also supported by qualitative findings (Section~\ref{qual:preference}). However, icons sometimes suggested a more balanced collaboration than intended, showing the need to refine the design to avoid misleading readers. Furthermore, this visualization may not easily scale toward representing multiple journalists and other editorial roles, as roles executed in parallel would disrupt the linear sequence and can then result in visual clutter. As Altay \& Gilardi \cite{altay2024people} note, distinctions between human and AI contributions must be very clear so users do not overestimate or underestimate AI’s role. 

\textbf{Chatbot} enabled in-depth exploration through progressive disclosure \cite{springer2020progressive} which gave users more control. It lead to the longest dwell times and highest fixation and saccade counts, indicating high engagement, but not necessarily more information processing per element. Quantitative findings showed it was the most informative, but also significantly harder to understand than all other visualizations, confirming interview comments that it could feel overwhelming (Section~\ref{qual:preference}). While valued in high-stake contexts, some found it offered too much to read, where its lack of immediate overview could additionally discourage casual news readers. This aligns with Burrus et al.'s \cite{burrus2024unmasking} perspective of cautioning against overloading users with technical detail. 

\textbf{Task-based Timeline} provided a good overview, insight into specific steps, and was most frequently preferred. Questionnaire and gaze data indicated effective information processing. However, the hover interaction was not always discovered, suggesting further refinement by replacing hover with inline text (Section~\ref{qual:preference}), again aligning with Burrus et al. \cite{burrus2024unmasking} to inform users without too much technical detail.

To summarize, we find that effectiveness depends on the intended use and context. Textual disclosure is familiar, and while increasingly pervasive across work practices \cite{Constantinides2025futureofworkblended}, is the least effective in communicating collaboration. Timelines provide strong overviews and insights into the specific steps of the process, where we believe this is generally suitable for human-AI collaboration tasks with (currently) clear workflows (e.g., producing an academic research manuscript \cite{Elagroudy2024HCICycle}, or practicing design work \cite{li2024user}). For high-stake articles, and more generally safety-critical domains (e.g., medical AI systems \cite{Moreno2026trustwothyai-healthcare}) that may require detailed information, the Chatbot would be most effective. The amplification effects raise the question: do disclosures simply inform readers, or do they also reframe how AI’s role is understood? Our findings show that design choices can tilt perceptions toward human or AI involvement, making disclosures active framings of contributions rather than neutral sources of information. This is in line with recent research that positions framing as a core dimension of design practice, actively shaping narratives, not just showing data \cite{shukla2025framing}. To that end, if we are to maximize both transparency and usability, disclosure formats should be adapted to a news article's topic, desired information depth, and the intended level of reader engagement. However, we caution about blanket personalization of disclosure visualizations and instead recommend consistency, specifically due to any risks of end-user misinterpretation especially should disclosures be used within sensitive settings (e.g., patient-centered health care \cite{Perlis2025healthcare_aidisclosure} and medical AI system design \cite{Moreno2026trustwothyai-healthcare}).



\subsection{Design considerations}

\textbf{Placement as a balance between visibility and bias.}  
Participants were divided on whether disclosures should appear at the beginning of articles to encourage critical reading or at the end to avoid introducing negative bias (Sec.~\ref{qual:realworld}). Prior studies show that AI-labeled headlines are perceived as less accurate or less shareable \cite{altay2024people,toff2024or,gilardi2024willingness}, reflecting the transparency dilemma where disclosures intended to increase trust instead end up eroding it \cite{SCHILKE2025104405trust}. Whether our human–AI collaboration visualizations, which are more detailed, influence users similarly remains an open question. However, it does raise the consideration of how to design disclosures so they are visible enough to support critical engagement without introducing unintended bias (cf., progressive disclosures \cite{springer2020progressive}), and whether other means such as toggles may be more suitable.  



\textbf{Disclosure formats and stake dependence.}  
Although our user study stimuli consisted of only low-stakes articles, our qualitative findings support \citet{gamage2025synthetic}'s findings, where we found that participants considered stake contexts in news articles, which often demanded different formats. For low-stakes articles, a simple textual label may suffice, even if less effective overall. While high-stakes or sensitive topics may require more detailed visualizations, such as timelines or a chatbot format (Section~\ref{qual:realworld}) to provide more nuance and accountability. News organizations could consider developing internal guidelines to match disclosure formats to article type and stake dependence, similar to editorial policies. 

\textbf{Supporting AI literacy through disclosures.}  
Several participants reported becoming more critical readers or more aware of AI’s capabilities after seeing the disclosures (Section~\ref{qual:realworld}). This suggests that disclosures may function as implicit AI literacy (cf., \cite{Velander2024ailiteracyconceptualizations}) interventions. Repeated exposure could increase literacy dimensions such as understanding, detection, and ethical awareness, in line with the commonly used Meta AI Literacy Scale (MAILS) \cite{carolus2023mails}. While not necessarily designed as educational tools, meaningful human-AI collaboration disclosure designs could have the added benefit of increasing AI literacy among audiences with limited GenAI knowledge.

\textbf{Disclosures for AI-generated disclosures.} 
Despite our detailed visualizations, some participants still questioned the truthfulness of the disclosures themselves \ref{qual:interpretation}. Given that LLMs can hallucinate or fabricate believable but incorrect information \cite{ji2023hallucination}, this raises concerns for any AI-generated disclosures, most prominently Chatbot-based disclosures. In this regard, AI-generated disclosures raise a new problem: what if the disclosure itself is either subject to AI hallucination or deliberately populated with incorrect information? Does this necessitate an additional disclosure on the AI-generated disclosure? Future work should explore how to prevent disclosures from becoming sources of misinformation by indicating how, by whom, and with what information they were created, and by investigating how trust indicators can aid users in evaluating their accuracy and trustworthiness.  

\textbf{Encoding disclosures into provenance signals}
AI usage detection approaches across modalities such as watermarking \cite{kirchenbauer2023watermark}, structural marking \cite{block2025gaussmark}, metadata standards \cite{Trattner2025}, or cryptographic signatures \cite{Pallav2025Cryptographic} are key steps toward detection of AI-generated content, but typically do not include human-AI collaboration ratios. For example, prior work on metadata standards like C2PA have shown that provenance labels can improve trust \cite{Trattner2025}, whereas others such as TransparentMeta \cite{TransparentMeta2025} aim to encrypt transparency and attribution metadata directly into audio files. Our findings open the possibility of extending provenance frameworks to include metadata on how humans and AI collaborated across tasks and roles, and to encode these disclosures as machine-readable provenance signals.

\subsection{Limitations and future work}\label{sec:limitation}
First, it can be argued that since we are concerned with the news production cycle, we should have adopted a more participatory AI approach (Cf., \cite{Birhane2022,Hansen2020} where we involved journalists directly, as this may limit insights into the feasibility of implementing such disclosures within editorial workflows. However, this was intentionally scoped as our study was primarily focused on establishing whether and how readers make sense of human-AI collaboration disclosure visualizations, before turning to questions of journalist adoption and/or their preferred methods of disclosure creation. Nevertheless, we believe this provides a rich avenue for future work: from establishing how journalists would like to disclose AI usage, to testing disclosure visualizations on live news platforms, and across articles spanning both low and high stakes, and with keeping the target demographic in mind (e.g., race or gender \cite{cheong2025penalizingtransparencyaidisclosure} or political orientation \cite{araujo2023humans}). Another cautionary aspect: While it is known that LLM users do engage in passive non-disclosure or active concealment of the usage of AI \cite{zhang2024secret}, this may not necessarily generalize to journalism, where trust and accountability are crucial. To that end, we contend that for true adoption, collaborations with journalists and news organizations will be essential to evaluate workflow integration and feasibility with respect to this emergent type of AI system disclosures. 


Third, our study was limited to evaluating only four prototype designs. While these were selected to create diverse visualizations and keeping the amount feasible for evaluation, other design ideas were generated which can be further explored. 
Additionally, all prototypes communicated the same core information about human and AI contributions, but differed in the level of detail provided. In particular, the Chatbot enabled users to retrieve more details through the layered predefined questions, incorporating progressive disclosure (cf., \cite{springer2020progressive}). This difference in information granularity was intentional to explore how readers respond to varying disclosure formats that naturally present different information depths. This choice ensured diversity in disclosure design, despite introducing a potential confound. Moreover, our disclosure designs were restricted primarily to textual news content, where we do not examine multimodal disclosures (e.g., in audio, images, or video). For example, image-based disclosures such as C2PA visual provenance approaches were not explored, where embedding C2PA provenance labels in images has shown to improve trust in AI-generated content \cite{Trattner2025}. 
Fourth, there are clear limits to our ecological validity; despite that our study was conducted with a controlled web application, and aimed to mimic web news consumption, this does not fully reflect the context of real-world news platforms, which also include branding that may influence trust \cite{liu2019machine}. Relatedly, our study only explored short-term effects of AI disclosures, where it remains future work to better understand the long-term impact of repeated exposures to such disclosures, where participants may get desensitized and stop noticing or reading them after the initial short-term engagement (cf., \cite{gilardi2024willingness}). 

Finally, given that users have different disclosure visualization preferences (cf., \cite{Venkatraj2025,scharowski2023certification}), it is worth asking to what extent media organizations should personalize these disclosures to end-users visiting their websites. While newsrooms can assess user preferences and tailor such interfaces, with a manageable workflow should such metadata tagging become standardized, the real risk lies in differing interpretations communicated, where AI roles and contributions can be more/less amplified depending on the disclosure visualization. This may pose risks to news organizations, especially given that prior work has shown that AI-generated news is perceived as less credible \cite{Longoni2022}, and differences in trust with some disclosure designs more than others \cite{gamage2025synthetic,scharowski2023certification}, even if limited to high stakes scenarios. Nevertheless, we speculate that while in the future users may have a minimal (standardized) disclosure visualization available for all, there could be optional expansion layers of tailored visualizations for readers that desire this. This would provide a combination of media transparency, user control, and importantly, editorial consistency.

\section{Conclusion}
This work aimed to examine how human–AI collaboration in news production can be communicated and perceived through visual disclosures. Based on co-design sessions (N=10), we derive four disclosure visualizations and evaluate them in a controlled within-subjects lab study (N=32). We examined how these disclosure visualizations (Textual, Role-based Timeline, Task-based Timeline, Chatbot) and collaboration ratios (Primarily Human-written vs. Primarily AI) influenced visualization perceptions, gaze patterns, and post-experience responses. We found that (a) all prototypes effectively communicated human-AI collaboration output ratios, but for varied purposes; (b) textual disclosures were least effective in communicating human-AI collaboration; (c) Role-based and Task-based Timelines provided clearer overviews of editorial steps, while the Chatbot offered the most in-depth information; and (d) role-based timelines amplified AI contribution in primarily human-written articles, whereas task-based timeline shifted perceptions toward human involvement in primarily AI articles. Our work provides empirically validated human-AI collaboration disclosure prototypes\footnote{Source code for all visualizations will be made publicly available after publication.} and design considerations that can support researchers, practitioners, and news organizations in creating more nuanced disclosures. Through such disclosures, we envision a future where human and AI contributions can be better captured and communicated, especially for critical domains such as journalistic editorial processes.

\section*{AI Usage Disclosure Statement}
\begin{figure}[h!]
    \centering
    \includegraphics[width=0.9\linewidth]{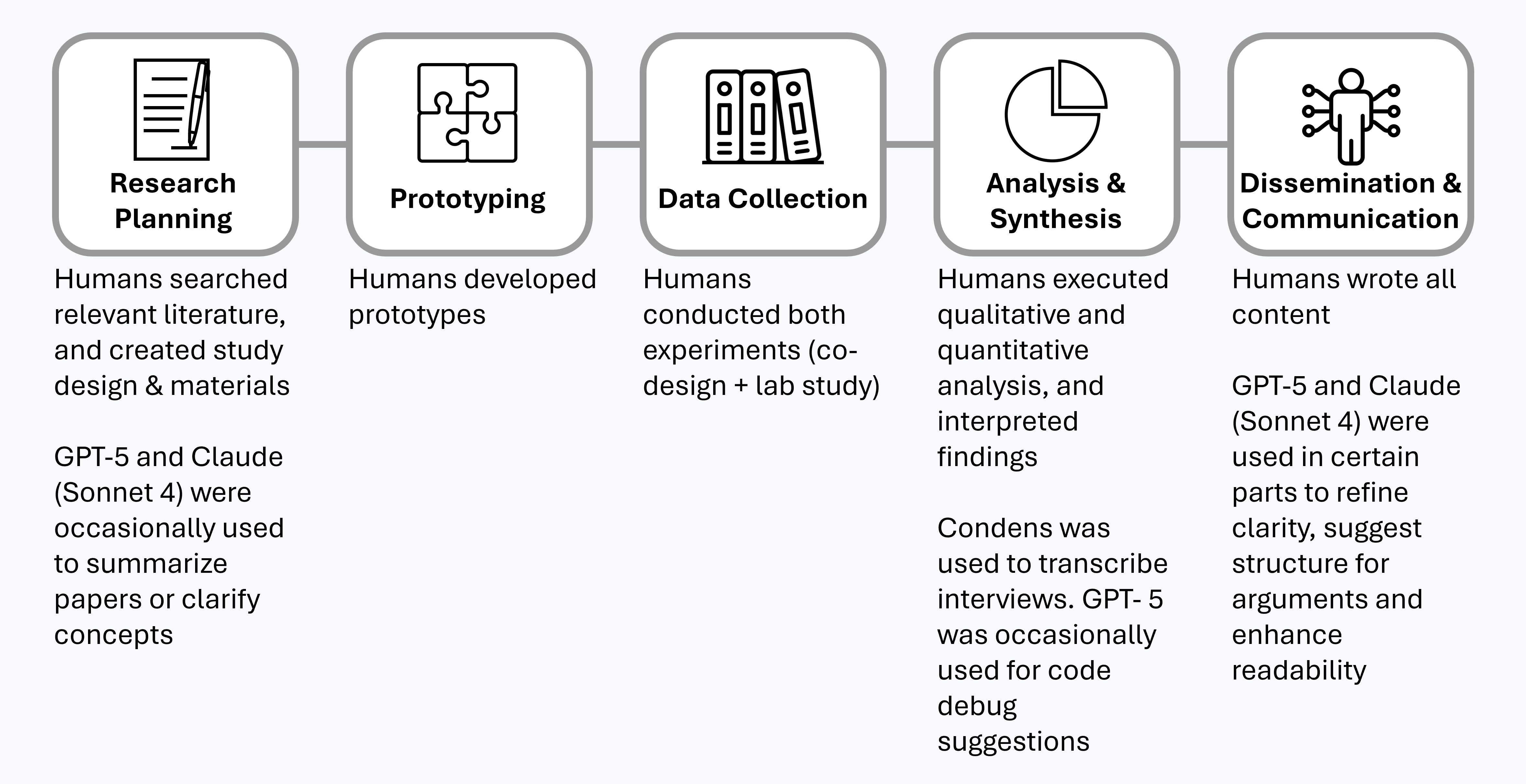}
    \caption{Task-based Timeline to disclose our AI usage.}
    \label{fig:disclstatement}
    \Description{Task-based Timeline showing five stages of the research workflow. Each stage is represented with an icon and text describing human and AI contributions. (1) Research planning: humans designed the study, with ChatGPT-5 and Claude (Sonnet 4) occasionally used to summarize papers and clarify concepts. (2) Prototyping and (3) data collection were entirely conducted by humans. (4) Analysis and synthesis were executed by humans, with Condens used for transcription and ChatGPT-5.0 for occasional debugging suggestions. (5) Dissemination and communication: humans wrote the paper, with limited AI use for refining clarity and readability.}
\end{figure}

To align with the disclosure practices we investigated, we conclude with a disclosure statement of our own collaboration with AI (see Fig~\ref{fig:disclstatement}). This illustrates how human-AI collaboration disclosure visualizations could also be adapted to research, with five research phases commonly used in HCI \cite{Elagroudy2024HCICycle}. We emphasize that the intellectual framing, the initial content created, the critical perspectives, and final interpretations and crafting of the discussion, remain our own. 

\bibliographystyle{ACM-Reference-Format}
\interlinepenalty=10000
\bibliography{disclosure_vis}

\end{document}